# Optical Communication in Space: Challenges and Mitigation Techniques

Hemani Kaushal[1] and Georges Kaddoum[2]

[1]Department of Electrical, Electronics and Communication Engineering, The NorthCap University, Gurgaon, Haryana, India-122017.

[2]Département de génie électrique, École de technologie supérieure, Montréal (QC), Canada.

*Abstract*—In recent years, free space optical (FSO) communication has gained significant importance owing to its unique features: large bandwidth, license free spectrum, high data rate, easy and quick deployability, less power and low mass requirements. FSO communication uses optical carrier in the near infrared (IR) band to establish either terrestrial links within the Earth's atmosphere or inter-satellite/deep space links or ground-to-satellite/satellite-to-ground links. It also finds its applications in remote sensing, radio astronomy, military, disaster recovery, last mile access, backhaul for wireless cellular networks and many more. However, despite of great potential of FSO communication, its performance is limited by the adverse effects (viz., absorption, scattering and turbulence) of the atmospheric channel. Out of these three effects, the atmospheric turbulence is a major challenge that may lead to serious degradation in the bit error rate (BER) performance of the system and make the communication link infeasible. This paper presents a comprehensive survey on various challenges faced by FSO communication system for ground-to-satellite/satellite-to-ground and inter-satellite links. It also provide details of various performance mitigation techniques in order to have high link availability and reliability. The first part of the paper will focus on various types of impairments that pose a serious challenge to the performance of optical communication system for ground-to-satellite/satellite-to-ground and inter-satellite links. The latter part of the paper will provide the reader with an exhaustive review of various techniques both at physical layer as well as at the other layers (link, network or transport layer) to combat the adverse effects of the atmosphere. It also uniquely presents a recently developed technique using orbital angular momentum for utilizing the high capacity advantage of optical carrier in case of space-based and near-Earth optical communication links. This survey provides the reader with comprehensive details on the use of space-based optical backhaul links in order to provide high capacity and low cost backhaul solutions.

## I. Introduction

### A. FSO Communication - An Overview

In the recent few years, tremendous growth and advancements have been observed in information and communication technologies. With the increasing usage of high speed internet, video-conferencing, live streaming etc., the bandwidth and capacity requirements are increasing drastically. This ever growing demand of increase in data and multimedia services has led to congestion in conventionally used radio frequency (RF) spectrum and arises a need to shift from RF carrier to optical carrier. Unlike RF carrier where spectrum usage is restricted, optical carrier does not require any spectrum licensing and therefore, is an attractive prospect for high bandwidth and capacity applications. Optical wireless communication (OWC) is the technology that uses optical carrier to transfer information from one point to another through an unguided channel which may be an atmosphere or free space. OWC is considered as a next frontier for high speed broadband connection as it offers extremely high bandwidth, ease of deployment, unlicensed spectrum allocation, reduced power consumption ($\sim$1/2 of RF), reduced size ($\sim$1/10 of the RF antenna diameter) and improved channel security [1]. It provides LOS communication owing to its narrow transmit beamwidth and works in visible and IR spectrum. The basic principle of OWC is similar to fiber optic communication except that unlike fiber transmission, in this case the modulated data is transmitted through unguided channel instead of guided optical fiber. The initial work on OWC has started almost 50 years back for defense and space applications where US military used to send telegraph signal from one point to another using sunlight powered devices. In year 1876, Alexander Graham Bell demonstrated his first wireless telephone system [2], [3] by converting sound waves to electrical telephone signals and transmitted the voice signal over few feets using sunlight as carrier. The device was called "photo-phone" as it was the world's first wireless telephone system. Thereafter, with the discovery of first working laser at Hughes Research Laboratories, Malibu, California in 1960 [4], a great advancement was observed in FSO technology.

The OWC can be classified into two broad categories, namely indoor and outdoor optical wireless communications. Indoor OWC uses IR or visible light for communicating within a building where the possibility of setting up a physical wired connection is cumbersome [5]–[12]. Indoor OWC is classified into four generic system configurations i.e., directed line-of-sight (LOS), non-directed LOS, diffused and tracked. Outdoor OWC is also termed as free space optical (FSO) communication. The FSO communication systems are further classified into terrestrial and space optical links that include building-to-building, ground-to-satellite, satellite-to-ground, satellite-to-satellite, satellite-to-airborne platforms (unmanned aerial vehicles (UAVs) or balloons), [13]–[15] etc. Fig. 1 illustrates the classification of OWC system. This survey is focused around space optical links which include both ground-to-satellite/ satellite-to-ground links, inter-satellite links and deep space links.

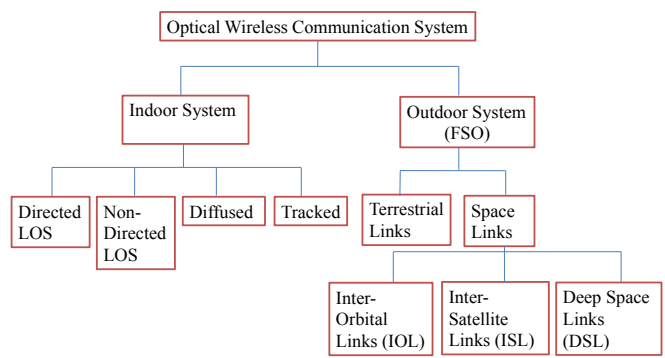

Figure 1. Classification of the optical wireless communication system

Advancement in space technology and development of more sophisticated space-based instruments opened a new chapter for optical space-based communication. Due to increasing demands for high data rate and large communication capacity, researchers are actively working to build all optical communication architecture that includes ground-to-satellite optical communication links which are connected to satellite optical network and satellite-to-ground optical links as shown in Fig. 2.

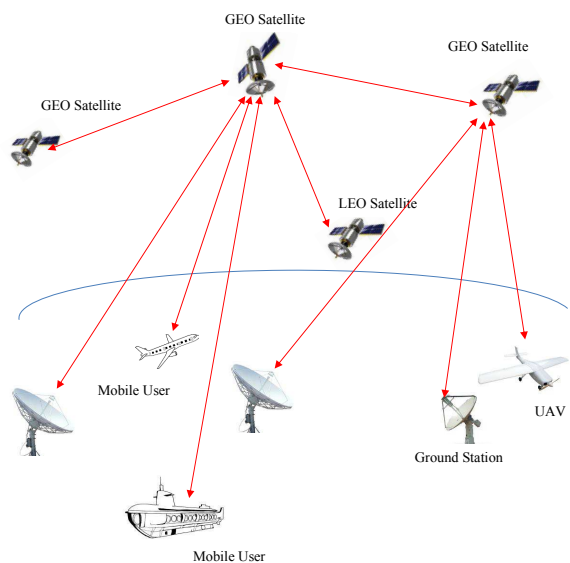

Figure 2. Space FSO links

The first theoretical study on optical uplink transmission from ground-to-satellite has been studied by Fried in 1967 [16]. Few years later, an uplink transmission using ground-based continuous-wave (CW) argon laser towards geodetic Earth orbiting satellite-II (GEOS-II) was demonstrated in [17]. Thereafter, various theoretical studies were suggested [14], [18]–[20] and successful experiments [21]–[23] were performed to investigate optical ground-to-satellite and inter-satellite communications. In early 1990, a relay mirror experiment (RME) was conducted using three laser beams that propagated from ground-to-satellite and were retro-reflected from the RME spacecraft orbiting at an altitude of 350 km [24]. The beam intensity profile was measured for investigating the temporal nature of atmospheric turbulence on the optical beam. In 1992, an uplink optical communication to deep space vehicle was demonstrated through Galileo optical experiment (GOPEX) that transmitted a pulsed laser signal from two optical ground stations (OGS) mounted at California and New Mexico [25]. The results demonstrated the distortion of uplink beam due to atmospheric turbulence. Later in 1995, the first ground-to-space two way communication link was demonstrated in ground/orbiter lasercom demonstration (GOLD) using argon ion laser [26], [27]. A comparison of theoretical and experimental data for single and multiple uplink beams was carried out in the GOLD demonstration. A bi-directional Earth-to-moon laser link was demonstrated with adaptive optics to mitigate the effect of atmospheric turbulence in [28]. The first inter-satellite laser communication link was successfully demonstrated by European Space Agency (ESA) between two satellites SPOT-4 and ARTEMIS for optical data-relay services at 50 Mbps [29]. They built an OGS and commission Semi-conductor Inter Satellite Link Experiment (SILEX) terminals in space. Later, successful bi-directional optical link between KIRARI, the Japanese satellite (officially called Optical Inter-Orbit Communications Engineering Test Satellite - OICETS) and ESA's Artemis was demonstrated in 2005 [30]. An optical link between two LEO orbiting satellites, Terra SAR-X and NFIRE, at 5.5 Gbps on a total distance of 5500 km and at a speed of 25, 000 km/hr has been established in 2008. The first successful ground-satellite optical link was conducted between the OGS and ETS-VI satellite in Konegi, Japan [31].

Several other experiments were performed in military and aerospace laboratories that demonstrated ground-to-satellite, satellite-to-satellite and satellite-to-ground optical links. It has resulted in various successful missions like (i) airborne flight test system (AFTS)- a link between aircraft and ground station at New Mexico [32], (ii) laser cross link system (LCLS)- full duplex space-to-space link for geosynchronous system [33], (iii) optical communication demonstrator (OCD)- laboratory prototype for demonstrating high speed data transfer from satellite-to-ground, (iv) stratospheric optical payload experiment STROPEX (CAPANINA Project)- high bit rate optical downlink from airborne station to transportable optical ground station [34], (v) Mars laser communications demonstration (MLCD)- provides up to 10 Mbps data transfer between Earth and Mars [35], and (vi) airborne laser optical link (LOLA)- first demonstration of a two-way optical link between high altitude aircraft and GEO satellite (ARTEMIS) [36]. Another mission by NASA is laser communication relay demonstration (LCRD) that will be launched in 2017 will demonstrate optical relay services for near earth and deep space communication missions [37]

This paper presents a comprehensive survey of FSO communication with primary focus on ground-to-satellite, satellite-to-ground and inter-satellite links. The issues involved in laser uplink are different from that of downlink or inter-satellite links. In case of laser uplink from ground-to-satellite, the beam comes immediately in contact with the atmosphere and therefore, suffer more from distortion and pointing instability due to spatial and temporal changes in the refractive index of the atmosphere. On the other hand, the downlink communication from satellite-to-ground causes the optical beam to spread geometrically (i.e., caused primarily due to beam divergence loss) and very little spread is due to atmospheric effect or variation in the beam steering. For this reason, the effect of atmospheric turbulence is generally very small on the downlink propagation as the beam goes through a non-atmospheric path until it reaches about 30 km from the Earth's surface. In case of inter-satellite links, the major challenge is laser beam pointing to or from a moving platform. For this reason, a very tight acquisition, tracking and pointing (ATP) systems are required for the optical beam to reliably reach the receiver. This survey paper provides an exhaustive coverage of various challenges and their mitigation techniques for space-based optical communication links.

### B. Advantages of FSO Communication over RF Communication

FSO communication system offers several advantages over RF system. The major difference between FSO and RF communication arises from the large difference in the wavelength. For FSO system, under clear weather conditions (visibility > 10 miles), the atmospheric transmission window lies in the near IR wavelength range between 700 nm to 1600 nm. The transmission window for RF system lies between 30 mm to 3 m. Therefore, RF wavelength is thousand of times larger than optical wavelength. This high ratio of wavelength leads to some interesting differences between the two systems as given below:

(I) **High bandwidth:** It is a well known fact that an increase in carrier frequency increases the information carrying capacity of a communication system. In RF and microwave communication systems, the allowable bandwidth can be up to 20% of the carrier frequency. In optical communication, even if the bandwidth is taken to be 1% of carrier frequency ($\approx 10^{16}$ Hz), the allowable bandwidth will be 100 THz. This makes the usable bandwidth at an optical frequency in the order of THz which is almost $10^5$ times that of a typical RF carrier [38], [39].

(II) **Less power and mass requirements:** The beam divergence is proportional to $\lambda/D_R$, where $\lambda$ is the carrier wavelength and $D_R$ the aperture diameter. Thus, the beam spread offered by the optical carrier is narrower than that of the RF carrier. This leads to an increase in the intensity of signal at the receiver for a given transmitted power. Fig. 3 shows the comparison of beam divergence for optical and RF signals when sent back from Mars towards Earth.

Thus, a smaller wavelength of optical carrier permits the FSO designer to come up with a system that has smaller antenna than RF system to achieve the same gain (as antenna gain scales inversely proportional to the square of operating wavelength). The typical size for the optical system is 0.3 m vs 1.5 m for RF spacecraft antenna [26]. Table I gives the power and mass comparison between optical and RF communication systems using 10 W and 50 W for optical and Ka band systems, respectively at 2.5 Gbps.

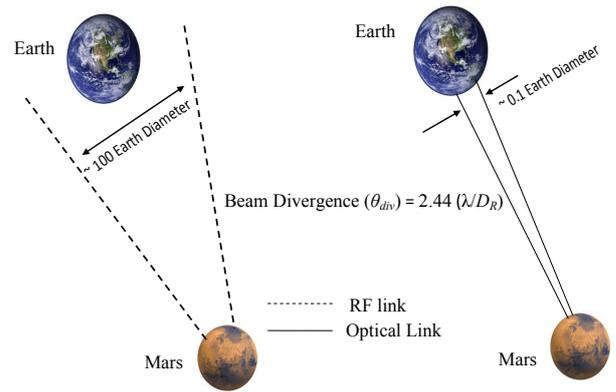

Figure 3. Comparison of optical and RF beam divergence from Mars towards Earth [40]

| Link | Optical | RF |
|---|---|---|
| **GEO-LEO** | | |
| Antenna Diameter | 10.2 cm (1.0) | 2.2 m (21.6) |
| Mass | 65.3 kg (1.0) | 152.8 kg (2.3) |
| Power | 93.8 W (1.0) | 213.9 W (2.3) |
| **GEO-GEO** | | |
| Antenna Diameter | 13.5 cm (1.0) | 2.1 m (15.6) |
| Mass | 86.4 kg (1.0) | 145.8 kg (1.7) |
| Power | 124.2 W (1.0) | 204.2 W (1.6) |
| **LEO-LEO** | | |
| Antenna Diameter | 3.6 cm (1.0) | 0.8 m (22.2) |
| Mass | 23.0 kg (1.0) | 55.6 kg (2.4) |
| Power | 33.1 W (1.0) | 77.8 W (2.3) |

Table I
COMPARISON OF POWER AND MASS FOR GEOSTATIONARY EARTH ORBIT (GEO) AND LOW EARTH ORBIT (LEO) LINKS USING OPTICAL AND RF COMMUNICATION SYSTEMS (VALUES IN PARENTHESES ARE NORMALIZED TO THE OPTICAL PARAMETERS) [41]

**High directivity:** The directivity of antenna is closely related to its gain. The advantage of optical carrier over RF carrier can be seen from the ratio of antenna gain as given in the equation below [42]

$$\frac{\text{Gain}_{(\text{optical})}}{\text{Gain}_{(\text{RF})}} \approx \frac{4\pi/\theta^2_{div(\text{optical})}}{4\pi/\theta^2_{div(\text{RF})}}, \quad (1)$$

where $\theta_{div(\text{optical})}$ and $\theta_{div(\text{RF})}$ are the optical and RF beam divergences, respectively and are proportional to $\lambda/D_R$. Since the optical wavelength is very small, a very high directivity and improved gain are obtained.

(III) (IV) **Unlicensed spectrum:** In the RF system, interference from adjacent carrier is the major problem

due to spectrum congestion. This requires the need of spectrum licensing by regulatory authorities. But on the other hand, the optical system is free from spectrum licensing till now. This reduces the initial set up cost and development time [43].

(V) **High Security:** FSO communication can not be detected by spectrum analyzers or RF meters as FSO laser beam is highly directional with very narrow beam divergence. Any kind of interception is therefore very difficult. Unlike RF signal, FSO signal cannot penetrate walls, therefore can prevent eavesdropping [44].

In addition to the above advantages, FSO communication offers secondary benefits as being : (i) easily expandable and reducing the size of network segments, (ii) light weight and compact, (iii) easy and quick deployable, and (iv) able to be used where fiber optic cables cannot be used. However, despite of many advantages, FSO communication system has its own drawbacks over RF system. The main disadvantage is the requirement of tight ATP system due to the narrow beam divergence. Also, FSO communication is dependent upon unpredictable atmospheric conditions that can degrade the performance of the system. Another limiting factor, is the position of the Sun relative to the laser transmitter and receiver. In a particular alignment, solar background radiations can increase and that will lead to poor system performance [45], [46]. This undoubtedly poses a great challenge to FSO system designers.

*C. Choice of wavelength in FSO communication*

Wavelength selection in FSO communication is a very important design parameter as it affects link performance and detector sensitivity of the system. Since antenna gain is inversely proportional to operating wavelength, therefore, it is more beneficial to operate at lower wavelengths. However, higher wavelengths provide better link quality and lower pointing-induced signal fades [47]. Therefore, a careful optimization of operating wavelength in the design of FSO link helps in achieving a better performance. The choice of wavelength strongly depends on atmospheric effects, attenuation and background noise power. Further, the availability of transmitter and receiver components, eye safety regulations and cost deeply impacts the selection of wavelength in the FSO design process.

The International Commission on Illumination [48] has classified optical radiations into three categories: IR-A (700 nm to 1400 nm), IR-B (1400 nm to 3000 nm) and IR-C (3000 nm to 1 mm). It can be sub-classified into : (i) near-infrared (NIR) ranging from 750 nm to 1450 nm is a low attenuation window and mainly used for fiber optics, (ii) short-infrared (SIR) ranging from 1400 nm to 3000 nm out of which 1530 nm to 1560 nm is a dominant spectral range for long distance communication, (iii) mid-infrared (MIR) ranging from 3000 nm to 8000 nm is used in military applications for guiding missiles, (iv) long-infrared (LIR) ranging from 8000 nm to 15 $\mu$m is used in thermal imaging, and (v) far-infrared (FIR) is ranging from 15 $\mu$m to 1 mm. Almost all commercially available FSO systems are using NIR and SIR wavelength ranges since these wavelengths are also used in fiber optic communication and their components are readily available in market.

For space-based optical applications, the choice of operating wavelength depends upon the trade-off between receiver sensitivity and pointing bias due to thermal variations across the Earth's surface. Generally, longer wavelengths are preferred as they cause reduction in solar background and solar scattering from the surface of the Earth. Lasers currently being considered and developed for space communication are in the range of 500 nm to 2000 nm. Table II summarizes various wavelengths used in practical space-based optical systems.

The wavelength selection for FSO communication has to be eye and skin safe as certain wavelengths between 400 nm to 1500 nm can cause potential eye hazards or damages to the retina [49]. Under International Electrotechnical Commission (IEC), lasers are classified into four groups from Class 1 to Class 4 depending upon their power and possible hazards [50]. Most of the FSO system use Class 1 and 1M lasers. For same safety class, FSO system operating at 1500 nm can transmit more than 10 times optical power than system operating at shorter operating wavelengths like 750 nm or 850 nm. It is because cornea, the outer layer of the eye absorb the energy of the light at 1550 nm and does not allow it to focus on retina. Maximum possible exposure (MPE) [51] specifies a certain laser power level up to which person can be exposed without any hazardous effect on eye or skin.

*D. Related Surveys*

Although, FSO communication has been studied in various literatures before, however most of these surveys are centered around terrestrial FSO links and very less surveys are available for space-based optical links. For example, a survey paper by Khalighi and Uysal [67] has elaborated various issues in FSO link according to communication theory prospective. They have presented different types of losses encountered in terrestrial FSO communication, details on FSO transceiver, channel coding, modulation and ways to mitigate fading effects of atmospheric turbulence. Similarly, Bloom et al. [68] have quantitatively covered various aspects that affect the performance of terrestrial FSO link - atmospheric attenuation, scintillation, alignment or building motion, solar interference and line-of-sight obstructions. Another survey by Demers et al. in [69] solely focused on FSO communication for next generation cellular networks. An introductory paper on terrestrial FSO communication by Ghassemlooy et al. [13] and Henniger et al. [43] provide an overview of various challenges faced in the design of FSO communication. In [70], the authors laid emphasis on deep space optical communication requirements and its future prospective. Similarly, in [71], the authors presented various trends and key initiatives used in deep space optical links. They have discussed the optical communication road-map for meeting future requirements and performance benefits in deep space optical links. Our survey is also related to space-based optical communication with focus on various challenges, current status and latest research trends in this field. In our work, we are intending to provide the reader with an exhaustive survey of various challenges faced in ground-to-satellite/satellite-to-ground and inter-satellite optical communication links. We have highlighted various performance mitigation techniques both at (i) physical layer, and (ii) other layers (link, network or transport). To the author's best knowledge most of the surveys till date have covered mitigation techniques only at physical layer. This paper uniquely presents atmospheric mitigation techniques at other layers aswell. The survey also provides the reader with an exhaustive literature on the use of FSO technology for high capacity backhaul links from high altitude platforms (HAPs) or satellite-based network. A recent approach for improving the quality and data returns from deep space and near Earth optical communication using orbital angular momentum (OAM) has been highlighted.

| Mission | Laser | Wavelength | Other parameters | Application |
|---------|-------|------------|------------------|-------------|
| Semiconductor Inter-satellite Link Experiment (SILEX) [52] | AlGaAs laser diode | 830 nm | 60 mW, 25 cm telescope size, 50 Mbps, 6 $\mu$rad divergence, direct detection | Inter-satellite communication |
| Ground/Orbiter Lasercomm Demonstration (GOLD) [27] | Argon-ion laser/GaAs laser | Uplink: 514.5 nm Downlink: 830 nm | 13 W, 0.6 m and 1.2m transmitter and receiver telescopes size, respectively, 1.024 Mbps, 20 $\mu$rad divergence | Ground-to-satellite link |
| RF Optical System for Aurora (ROSA) [53] | Diode pumped Nd:YVO4 laser | 1064 nm | 6 W, 0.135 m and 10 m transmitter and receiver telescopes size, respectively, 320 kbps | Deep space missions |
| Deep Space Optical Link Communications Experiment (DOLCE) [54] | Master oscillator power amplifier (MOPA) | 1058 nm | 1 W, 10-20 Mbps | Inter-satellite/deep space missions |
| Mars Orbiter Laser Altimeter (MOLA) [55] | Diode pumped $Q$ switched Cr:Nd:YAG | 1064 nm | 32.4 W, 420 $\mu$rad divergence, 10 Hz pulse rate, 618 bps, 850 $\mu$rad receiver field-of-view (FOV) | Altimetry |
| General Atomics Aeronautical Systems, Inc. (GA-ASI) & TESAT [56] | Nd:YAG | 1064 nm | 2.6 Gbps | Remotely piloted aircraft (RPA) to LEO |
| Altair UAV-to-ground Lasercomm Demonstration [57] | Laser diode | 1550 nm | 200 mW, 2.5 Gbps, 19.5 $\mu$rad jitter error, 10 cm and 1 m uplink and downlink telescopes size, respectively | UAV-to-ground link |
| Mars Polar Lander [58] | AlGaAs laser diode | 880 nm | 400 nJ energy in 100 nsec pulses, 2.5 kHz rate, 128 kbps | Spectroscopy |
| Cloud-Aerosol Lidar and Infrared Pathfinder Satellite Observation (CALIPSO) [59] | Nd:YAG | 532 nm/1064 nm | 115 mJ energy, 20 Hz rate, 24 ns pulse | Altimetry |
| KIrari's Optical Downlink to Oberpfaffenhofen (KIODO) [60] | AlGaAs laser diode | 847 nm/810 nm | 50 Mbps, 40 cm and 4 m transmitter and receiver telescopes size, respectively, 5$\mu$rad divergence | Satellite-to-ground downlink |
| Airborne Laser Optical Link (LOLA) [36] | Lumics fiber laser diode | 800 nm | 300 mW, 50 Mbps | Aircraft and GEO satellite link |
| Tropospheric Emission Spectrometer (TES) [61] | Nd:YAG | 1064 nm | 360 W, 5 cm telescope size, 6.2 Mbps | Interferometry |
| Galileo Optical Experiment (GOPEX) [25] | Nd:YAG | 532 nm | 250 mJ, 12 ns pulse width, 110 $\mu$rad divergence, 0.6 m primary and 0.2 m secondary transmitter telescope size, 12.19 x 12.19 mm charge coupled device (CCD) array receiver | Deep space missions |
| Engineering Test Satellite VI (ETS-VI) [62] | AlGaAs laser diode (downlink) Argon laser (uplink) | Uplink: 510 nm Downlink: 830 nm | 13.8 mW, 1.024 Mbps bidirectional link, direct detection, 7.5 cm spacecraft telescope size, 1.5 m Earth station telescope | Bi-directional ground-to-satellite link |
| Optical Inter-orbit Communications Engineering Test Satellite (OICETS) [63] | Laser Diode | 819 nm | 200 mW, 2.048 Mbps, direct detection, 25 cm telescope size | Bi-directional Inter-orbit link |
| Solid State Laser Communications in Space (SOLACOS) [64] | Diode pumped Nd:YAG | 1064 nm | 1 W, 650 Mbps return channel and 10 Mbps forward channel, 15 cm telescope size, coherent reception | GEO-GEO link |
| Short Range Optical Inter-satellite Link (SROIL) [65] | Diode pumped Nd:YAG | 1064 nm | 40 W, 1.2 Gbps, 4 cm telescope size, BPSK homodyne detection | Inter-satellite link |
| Mars Laser Communications Demonstration (MLCD) [66] | Fiber laser | 1064 nm and 1076 nm | 5 W, 1- 30 Mbps, 30 cm transmitter telescope size and 5 m and 1.6 m receiver telescope size, 64 PPM | Deep space missions |

Table II
WAVELENGTHS USED IN PRACTICAL FSO COMMUNICATION SYSTEMS

*E. Paper Organization*

The rest of the paper is organized as follows: Section II describes major challenges faced by laser uplink/downlink and space-based optical links. Due to narrow beam divergence of the optical carrier, accurate acquisition and pointing becomes very challenging for the space-based optical communication links. Section III provides an exhaustive discussion on ATP for space-based optical links. Section IV presents various techniques both at physical and other layers (link, network or transport layer) to mitigate the adverse effects of the atmosphere. This section will also give the details of the hybrid RF/FSO system that provides a practical solution by backing up the FSO link with low data rate RF link. Section V highlights a recent study on a OAM-based FSO system used in combination with channel coding, MIMO or adaptive optics for enhancing data returns from deep space or near Earth communication in turbulent atmosphere. Section VI presents the use of FSO technology in space-based optical backhaul links. Section VII discusses the future scope of FSO technology and finally, the last section will conclude the survey paper.

## II. CHALLENGES IN SPACE-BASED OPTICAL COMMUNICATION

The ground-to-satellite and satellite-to-ground FSO communications are subject to atmospheric effects. FSO technology uses atmospheric channel as a propagating medium whose properties are random function of space and time. It makes FSO communication a random phenomena that is dependent on weather and geographical location. Various unpredictable environmental factors like clouds, snow, fog, rain, haze, etc., cause strong attenuation in the optical signal and limit the link distance. For inter-satellite FSO links, various limiting factors include pointing, background noise and link availability. This section will cover various challenges faced by system designers in laser uplink/downlink as well as inter-satellite FSO links.

*A. FSO Uplink/Downlink*

FSO signal from ground-to-satellite or satellite-to-ground requires the passage of signal through the atmosphere. The losses encountered during FSO uplink are very large compared to downlink as the beam begins to spread and accumulate distortion the very moment optical signal is emitted from ground-based terminal. In case of uplink, the source of disturbance is close to the source and therefore, it corresponds to spherical wave model. On the other hand, during downlink, the source of the disturbance is close to the receiver terminal and therefore, it corresponds to plane wave model. When a laser beam propagates through atmosphere, it experiences power loss due to various factors and a role of system design engineer is to carefully examine the system design requirements in order to combat with the random changes in the atmosphere. Changes in temperature and pressure with altitude influence the optical beam propagation in different ways. The mean temperature of atmosphere as a function of altitude is given in [72]. Also, the concentration profile of various atmospheric constituents and aerosol particles vary with altitude with largest concentration is up to 1 - 2 km immediately above the Earth's surface. Therefore, the transmittance spectrum of the atmosphere experienced during uplink at a given zenith angle and wavelength gives rise to forbidden bands where the atmospheric attenuation is very high. MODTRAN 4.0 [73], [74] simulations give the measure of atmospheric transmittance and background sky radiance for various altitude, aerosols distribution, etc. For a reliable FSO communication, the system design engineer needs to have thorough understanding of beam propagation through random atmosphere and its associated losses. Various losses encountered by the optical beam when propagating through the atmospheric optical channel are:

(I) **Absorption and scattering loss:** When the laser beam propagates through the Earth's atmosphere it may have to interact with various gas molecules and aerosols particles present in the atmosphere. The loss in the atmospheric channel is mainly due to absorption and scattering process and it is described by Beer's law [75]. At visible and IR wavelengths, the principal atmospheric absorbers are the molecules of water, carbon-dioxide and ozone [76], [77]. The atmospheric absorption is a wavelength-dependent phenomenon. Some typical values of molecular absorption coefficients are given in Table III for clear weather conditions. The wavelength range of the FSO communication system is chosen to have minimal absorption. This is referred to as atmospheric transmission window. In this window, the attenuation due to molecular or aerosol absorption is less than 0.2 dB/km. There are several transmission windows within the range of 700 - 1600 nm. Majority of FSO systems are designed to operate in the windows of 780 - 850 nm and 1520 - 1600 nm. These wavelengths have been chosen because of the readily availability of the transmitter and detector components at these wavelengths. The wavelength dependance of attenuation under different weather conditions is commonly available in databases like MORTRAN [78], LOWTRAN [79] and HITRAN.

Scattering of light is also responsible for degrading the performance of the FSO system. Like absorption, scattering is also strongly wavelength dependent. If the size of atmospheric particles are small in comparison with the optical wavelength, then Rayleigh scattering is produced. This scattering is quite prominent for FSO communication around visible or ultraviolet range i.e., wavelengths below 1 $\mu m$. However, it can be neglected at longer wavelengths near IR range. Particles like air molecules and haze cause Rayleigh scattering [81]. If the atmospheric particles size are comparable with the optical wavelength, then Mie scattering is produced. It is dominant near IR wavelength range or longer. Aerosol particles, fog and haze are major contributors of Mie scattering. If the atmospheric particles are much larger than the optical wavelength like in case of rain, snow and hail, then the scattering is better described by geometrical optic models [82], [83].

| S.No | Wavelength (nm) | Molecular Absorption (dB/km) |
|---|---|---|
| 1. | 550 | 0.13 |
| 2. | 690 | 0.01 |
| 3. | 850 | 0.41 |
| 4. | 1550 | 0.01 |

Table III
MOLECULAR ABSORPTION AT TYPICAL WAVELENGTHS [80]

| Ground Station | Altitude | Horizon Angle | Transmittance | | |
|---|---|---|---|---|---|
| | | | 800 nm | 1060 nm | 1550 nm |
| Chang chun | 211 m | 21° | 0.15 | 0.25 | 0.43 |
| Kunming | 1899 m | 49° | 0.63 | 0.70 | 0.73 |
| Hainan | 20 m | 45° | 0.34 | 0.48 | 0.60 |
| Uru muqi | 846 m | 39° | 0.40 | 0.60 | 0.65 |
| Ali | 5022 m | 57° | 0.91 | 0.93 | 0.92 |

Table IV
MODTRAN SIMULATED ATMOSPHERIC TRANSMITTANCE FOR VARIOUS WAVELENGTH AT 5 KM OF VISUAL RANGE [84]

Total atmospheric attenuation is represented by the atmospheric attenuation coefficient, $\gamma$ which is expressed as a combination of absorption and scattering of light. It is therefore expressed as sum of four individual parameters given as

$$\gamma = \alpha_m + \alpha_a + \beta_m + \beta_a, \quad (2)$$

where $\alpha_m$ and $\alpha_a$ are molecular and aerosol absorption coefficients, respectively and $\beta_m$ and $\beta_a$ are molecular and aerosol scattering coefficients, respectively. It is to be mentioned that ground-to-satellite optical transmission at different altitudes above the sea level experiences variety of transmittance spectrum. Atmospheric transmittance is increased at high elevation angles as the impact due to aerosols particles is reduced. The atmospheric transmittance at a zenith angle, $\theta$ is expressed as [84]

$$T_{atm}(\lambda) = exp\left[-\sec\theta \int_0^H \gamma(\lambda, h)\, dh\right] \quad (3)$$

where $\gamma$ is the atmospheric attenuation coefficient, $H$ is the vertical height of the atmospheric channel, and $\lambda$ is the operating wavelength. An improvement in atmospheric transmittance is observed from 0.85 at sea level to 0.96 at 3 km at operating wavelength of 1000 nm with zero zenith angle [85]. Table IV shows the MODTRAN simulated atmospheric transmittance at 5 km visibility caused by longitude 77° GEO satellite and 5 ground stations located at mid-latitude in China continent. [84]

In [86], a rainfall model is developed considering various atmospheric conditions such as rainfall rate, attenuation, temperature and relative humidity. It also determines FSO scattering effect due to different rain drop sizes and rainfall velocity. Attenuation due to scattering caused by haze, fog and low cloud droplets was simulated in [87] and it was inferred that the size of the atmospheric droplets and their distribution plays a vital role in the FSO attenuation due to scattering. A simplified approach for FSO channel modeling is developed in [88] that uses the statistics of received signal power.

Various factors that cause absorption and scattering in FSO system are as follows:

- Fog: The major contribution for atmospheric attenuation is due to fog as it results in both absorption and scattering. During dense fog conditions, when the visibility is even less than 50 m, attenuation can be more than 350 dB/km [89]. This clearly shows that it could limit the availability of FSO link. In such cases, very high power lasers with special mitigation techniques help to improve the chances of link availability. Generally, 1550 nm lasers are a preferred choice during heavy attenuation because of their high transmitted power. Fog can extend vertically up to the height of 400 m above the Earth's surface. A comparison of fog attenuation properties for different operating wavelengths is studied in [90]. The attenuation due to fog can be predicted by applying Mie scattering theory. However, it involves complex computations and requires detailed information of fog parameters. Alternate approach is based on visibility range information, in which the attenuation due to fog is predicted using common empirical models. The wavelength of 550 nm is usually taken as the visibility range reference wavelength. Eq. (4) defines the specific attenuation of fog given by common empirical model for Mie scattering.

$$\beta_{fog}(\lambda) = \frac{3.91}{V}\left(\frac{\lambda}{550}\right)^{-p}, \quad (4)$$

where $V$(km) stands for visibility range, $\lambda$(nm) is the operating wavelength and $p$ the size distribution coefficient of scattering. The value of $p$ can be determined using Kim or Kruse model [91], [92]. Al Naboulsi et al. in [93] have provided relations to predict fog attenuation by characterizing advection and radiation fog, separately.
- Rain: The impact of rain is not much pronounced like that fog as rain droplets are significantly larger (100 to 10,000 $\mu m$) in size than the wavelength used in FSO communication. The attenuation loss for light rain (2.5 mm/hr) to heavy rain (25 mm/hr) ranges from 1 dB/km to 10 dB/km for wavelengths around 850 nm and 1500 nm [94], [95]. For this reason, the choice of hybrid RF/FSO systems improve the link availability especially for systems operating at 10 GHz frequency and above. This topic is discussed in more details in Sec. IV.

  The modeling of rain attenuation prediction is done using empirical methods proposed by International Telecommunication Union- Radio communication sector (ITU-R) for FSO communication [96]. The specific attenuation, $\alpha_{rain}$ (in dB/km) for a FSO link is given by [97]

  $$\alpha_{\text{rain}} = k_1 \mathbf{R}^{k_2}, \quad (5)$$

  where $\mathbf{R}$ is the rain rate in mm/hr and $k_1$ and $k_2$ are the model parameters whose values depend upon rain drop size and rain temperature. It is to be noted that rain accompanied by low clouds results in a very high attenuation. In order to combat for huge power loss during heavy rain accompanied by low clouds, high power lasers should be used and sufficient link margin greater than 30 dB should be achieved for maximum link availability of FSO system. Further, HAP interconnected optical links help in eliminating cloud blockage problem at different geographical locations [98]. Also, system designers prefer to adopt adaptive coding and modulation technique due to rapid variation in the atmospheric conditions between satellite and ground station [99].

- Snow: The size of snow particles are between fog and rain particles. Therefore, attenuation due to snow is more than rain but less than fog. During heavy snow, the path of the laser beam is blocked due to increasing density of snow flakes in the propagation path or due to the formation of ice on window pane. In this case, its attenuation is comparable to fog ranging between 30-350 dB/km and this can significantly reduce the link availability of the FSO system. For snow, attenuation is classified into dry and wet snow attenuation. The specific attenuation (dB/km), $\alpha_{snow}$ for snow rate $S$ in mm/hr is given as [96]

  $$\alpha_{\text{snow}} = aS^b, \quad (6)$$

  where the values of parameters $a$ and $b$ in dry and wet snow are

  $$\begin{array}{ll} \text{Dry snow}: & a = 5.42 \times 10^{-5} + 5.49, \quad b = 1.38, \\ \text{Wet snow}: & a = 1.02 \times 10^{-4} + 3.78, \quad b = 0.72. \end{array} \quad (7)$$

(II) **Atmospheric-turbulence:** Atmospheric turbulence is a random phenomenon which is caused by variation of temperature and pressure of the atmosphere along the propagation path. It will result in the formation of turbulent cells, also called eddies of different sizes and of different refractive indices. These eddies will act like a prism or lenses and will eventually cause constructive or destructive interference of the propagating beam. These perturbations in the wavefront introduced by the atmosphere can be physically described by the Kolmogorov model [100]. Depending on the size of turbulent eddies and transmitter beam size, three types of atmospheric turbulence effects can be identified:

- Turbulence-induced beam wander: Beam wander is a phenomenon which is experienced *when the size of turbulence eddies are larger than the beam size*. It will result in random deflection of the beam from its propagating path and leads to link failure. Beam wander is a major concern in case of uplink as the beam size is often smaller than the turbulence eddy side and it will result in the beam displacement by several hundred meters. The rms beam wander displacement is given as [101]

  $$\sigma_{BW}^2 \approx 0.54 (H - h_0)^2 \sec^2(\theta) \left(\frac{\lambda}{2W_0}\right)^2 \left(\frac{2W_0}{r_0}\right)^{5/3}, \quad (8)$$

  where $H$ and $h_0$ are the altitude of satellite and transmitter, respectively, $r_0$ is the atmospheric coherence length (also called Fried parameter) [18], $\lambda$ is the operating wavelength, $\theta$ is the zenith angle, $W_o$ is the initial beam size.

- Turbulence-induced beam spreading: Beam spreading takes into account *when the size of the eddies are smaller than the beam size*. In this case, the incoming beam will be diffracted and scattered independently leading to distortion of the received wavefront.

- Turbulence-induced beam scintillation: *When the eddy's size is of the order of beam size*, then the eddies will act like lens that will focus and de-focus the incoming beam. This will lead to redistribution of signal energy resulting in temporal and spatial irradiance fluctuations of the received signal. These intensity fluctuations of the received signal is known as scintillation and is the major cause of degradation in the performance of the FSO system.

Atmospheric turbulence also leads to the loss of spatial coherence of initially coherent beam. Structure function for phase fluctuations and correlation in the phase between two position vectors ($\rho_1$ and $\rho_2$) in the plane of observation across the beam is studied in [102], [103]. The degree of coherence for plane and spherical wave is expressed as

$$\gamma = exp\left[-\left(\frac{|\rho_1 - \rho_2|}{\rho_0}\right)^{5/3}\right] \quad (9)$$

where $\rho_0$ is the phase coherence radius. For $\rho > \rho_0$, the random phase angle is larger than $\pi$, and the wavefront looses its spatial coherence. Atmospheric turbulence may also produce depolarization of the light and temporal stretching of the optical pulse. Depolarized light causes significant reduction in the average power. In [104], an average power loss of about 160 dB is observed for depolarized light as compared to incident beam. The root-mean-square change in the polarization angle of an optical beam propagating through turbulent atmosphere is of the order of 10 - 9 rad/km which further increases linearly with the path length. The temporal spreading is due to multiple path lengths as the optical beam travels through the turbulent atmosphere. It severely limits the achievable bandwidth and the data rate. During the downlink of optical signal from satellite-to-ground, the main contribution for degradation of the signal quality is due to beam spreading, scintillation and loss of spatial coherence. For the uplink, beam wander and fluctuations in the angle of arrival at the receiver plane are the principle contributor for signal degradation.

Atmospheric scintillation is measured in terms of scintillation index (normalized variance of intensity fluctuations), $\sigma_I^2$ given by [100], [105]–[107]

$$\sigma_I^2 = \frac{\langle I^2 \rangle - \langle I \rangle^2}{\langle I \rangle^2} = \frac{\langle I^2 \rangle}{\langle I \rangle^2} - 1, \quad (10)$$

where $I$ is the irradiance (intensity) at some point in the detector plane and the angle brackets $\langle \rangle$ denotes an ensemble average. Scintillation index is expressed as variance of log-amplitude, $\sigma_x^2$ as

$$\sigma_I^2 \approx 4\sigma_x^2, \text{ for } \sigma_x^2 << 1. \quad (11)$$

Scintillation index is a function of the refractive index structure parameter, $C_n^2$. This parameter determines the strength of turbulence in the atmosphere. Clearly, $C_n^2$ will vary with time of day, geographical location and height. For near ground horizontal link, value of $C_n^2$ is almost constant and its typical value in case of weak turbulence is $10^{-17}$ m$^{-3/2}$ and for strong turbulence it can be up to $10^{-13}$ m$^{-3/2}$ or greater. For vertical links, the value of $C_n^2$ changes with altitude $h$ unlike horizontal link where its value is assumed to be constant. With the increase in the altitude, the value of $C_n^2$ decreases at the rate of $h^{-4/3}$. Therefore for vertical links, the value of $C_n^2$ has to be integrated over the complete propagation path extending from height of the receiver above sea level to the top of the atmosphere (roughly up to 40 kms). Due to this reason, the effect of atmospheric turbulence from ground-to-satellite (uplink) is different from satellite-to-ground (downlink). Various empirical models of $C_n^2$ have been proposed to estimate the turbulence profiles that are based on experimental measurements carried out at variety of geographical locations, times of day, wind speeds, terrain types, etc [108]. Some of the commonly used models are presented in Table V.

| Models | Range | Comments |
|---|---|---|
| PAMELA Model [109] | Long (few tens of kms) | - Robust model for different terrains and weather types<br>- Sensitive to wind speed<br>- Does not perform well over marine/overseas environment |
| NSLOT Model [110] | Long (few tens of kms) | - More accurate model for marine propagation<br>- Surface roughness is 'hard-wired' in this model<br>- Temperature inversion i.e., $(T_{air} - T_{sur} > 0)$ is problematic |
| Fried Model [111] | Short (in meters) | - Support weak, strong and moderate turbulence |
| Hufnagel and Stanley Model [112] | Long (few tens of kms) | - $C_n^2$ is proportional to $h^{-1}$<br>- Not suitable for various site conditions |
| Hufnagel Valley Model [113]–[115] | Long (few tens of kms) | - Most popular model as it allows easy variation of day-time and night-time profile by varying various site parameters like wind speed, iso-planatic angle and altitude<br>- Best suited for ground-to-satellite uplink<br>- HV 5/7 is a generally used to describe $C_n^2$ profile during day-time. HV5/7 yields a coherence length of 5 cm and isoplanatic angle of 7$\mu$rad at 0.5 $\mu$m wavelength |
| Gurvich Model [116] | Long (few tens of kms) | - Covers all regimes of turbulence from weak, moderate to strong<br>- $C_n^2$ dependance on altitude, $h$, follows power law i.e., $C_n^2 \propto h^{-n}$ where $n$ could be 4/3, 2/3 or 0 for unstable, neutral or stable atmospheric conditions, respectively. |
| Von Karman-Tatarski Model [117], [118] | Medium (few kms) | - Make use of phase peturbations of laser beam to estimate inner and outer scale of turbulence<br>- Sensitive to change in temperature difference |
| Greenwood Model [42], [119] | Long (few tens of kms) | - Night-time turbulence model for astronomical imaging from mountaintop sites |
| Submarine Laser Communication (SLC) [120] Model | Long (few tens of kms) | - Well suited for day-time turbulence profile at inland sites<br>- Developed for AMOS observatory in Maui, Hawaii |
| Clear 1 [121] | Long (few tens of kms) | - Well suited for night-time turbulent profile<br>- Averages and statistically interpolate radiosonde observation measurements obtained from large number of meteorological conditions |
| Aeronomy Laboratory Model (ALM) [122] | Long (few tens of kms) | - Shows good agreement with radar measurements<br>- Based on relationship proposed by Tatarski [118] and works well with radiosonde data |
| AFRL Radiosonde Model [123] | Long (few tens of kms) | - Similar to ALM but with simpler construction and more accurate results as two seperate models are used for troposphere and stratosphere<br>- Day-time measurements could give erroneous results due to solar heating of thermosonde probes |

Table V
TURBULENCE PROFILE MODELS FOR $C_n^2$

The most widely used model for vertical link is Hufnagel Valley Boundary (HVB) model [124] given by

$$\begin{aligned}C_n^2(h) &= 0.00594\left[\left(\frac{V}{27}\right)^2 \left(10^{-5}h\right)^{10}\exp\left(-h/1000\right)\right.\\ &\quad +2.7\times 10^{-16}\exp\left(-\frac{h}{1500}\right)\\ &\quad \left.+A\exp\left(-\frac{h}{100}\right)\right] m^{-2/3},\end{aligned} \quad (12)$$

where $V^2$ is the mean square value of the wind speed in m/s, $h$ is the altitude in meters and $A$ is a parameter whose value can be adjusted to fit various site conditions. The parameter $A$ is given as

$$\begin{aligned}A &= 1.29\times 10^{-12}r_0^{-5/3}\lambda^2 - 1.61\times 10^{-13}\theta_0^{-5/3}\lambda^2\\ &\quad +3.89\times 10^{-15}.\end{aligned} \quad (13)$$

In the above equation, $\theta_0$ is the isoplanatic angle [115] (angular distance over which the atmospheric turbulence is essentially unchanged) and $r_0$ is the atmospheric coherence length. The coherence length of the atmosphere is an important parameter that is dependent upon operating wavelength, $C_n^2$ and zenith angle $\theta$. For plane wave (downlink) propagating from altitude $h_o$ to $(h_o + L)$, it is given as

$$r_0 = \left[0.423k^2\sec(\theta)\int_{h_0}^{h_o+L} C_n^2(h)\,dh\right]^{-3/5}. \quad (14)$$

For spherical wave (uplink), it is expressed as [121]

$$r_0 = \left[0.423k^2\sec(\theta)\int_{h_0}^{h_o+L} C_n^2(h)\left\{\frac{L+h_o-h}{L}\right\}^{5/3}dh\right]^{-3/5}. \quad (15)$$

It is clear from above expressions that $r_0$ varies as $\lambda^{6/5}$, therefore, FSO link operating at higher wavelengths will have less impact of turbulence than at lower wavelengths. For the uplink, if transmitter beam size $W_0$ is of the order of $r_0$, a significant beam wander takes place. For downlink, the angle of arrival fluctuation [100] increases as the value of $r_0$ decreases. In case of weak turbulence, scintillation index for plane wave (downlink) can be written in terms of refractive index structure parameter, $C_n^2$ as

$$\sigma_I^2 = \sigma_R^2 \approx 2.24k^{7/6}\left(\sec(\theta)\right)^{11/6}\int_{h_0}^{h_o+L}C_n^2(h)h^{5/6}dh. \quad (16)$$

where $k$ is wave number $(2\pi/\lambda)$, $\sigma_R^2$ the Rytov variance, $L$ the link distance. It should be noted that weak fluctuation theory does not hold for larger zenith angles and smaller wavelengths. In that case, scintillation index for moderate to strong turbulence holds well and is given by [125]

$$\sigma_I^2 = \exp\left[\frac{0.49\sigma_R^2}{\left(1+1.11\sigma_R^{12/5}\right)^{7/6}} + \frac{0.51\sigma_R^2}{\left(1+0.69\sigma_R^{12/5}\right)^{7/6}}\right]-1. \quad (17)$$

For a weak turbulence i.e., $\sigma_I^2 < 1$, intensity statistics is given by the log-normal distribution. For a strong turbulence, $\sigma_I^2 \geq 1$, the field amplitude is Rayleigh-distributed which means negative exponential statistics for the intensity [111]. Besides these two models, a number of other statistical models [126] are used in literature to describe the scintillation statistics in either a regime of strong turbulence ($K$ model) or all the regimes ($I$-$K$, $M$ and Gamma-Gamma [127] models). For $3 < \sigma_I^2 < 4$, the intensity statistics is given by $K$ distribution. Another generalized form of $K$ distribution that is applicable to all conditions of atmospheric turbulence is $I$-$K$ distribution. However, $I$-$K$ distribution is difficult to express in closed form expressions. Similarly, log-normal Rician distribution is a generalized distribution that covers both weak and strong turbulences. It is applied if the optical field after propagating through turbulent atmosphere obeys Rice-Nakagami statistics with log-normal modulation factor. However, its numerical calculations in a closed loop solution are little cumbersome. In that case, the Gamma-Gamma distribution is used to successfully describe the scintillation statistics for weak to strong turbulence [128], [129]. In this model, the normalized irradiance $I$, is defined as the product of two independent random variables, i.e., $I = I_X I_Y$, where $I_X$ and $I_Y$ represents large-scale and small-scale turbulent eddies, respectively and each of them is governed by independent Gamma distribution. This leads to Gamma-Gamma distribution of $I$ whose probability density function (pdf) is given as

$$f_I(I) = \frac{2(\alpha\beta)^{(\alpha+\beta)/2}}{\Gamma(\alpha)\Gamma(\beta)}I^{((\alpha+\beta)/2)-1}K_{\alpha-\beta}\left(2\sqrt{\alpha\beta I}\right),\; I>0, \quad (18)$$

where $K_a(\cdot)$ is the modified Bessel function of second kind of order $a$. The parameters $\alpha$ and $\beta$ are the effective number of small scale and large scale eddies of the scattering environment. The scintillation index for this model is defined as

$$\sigma_I^2 = \frac{E\left[I^2\right]}{(E\left[I\right])^2} - 1 = \frac{1}{\alpha} + \frac{1}{\beta} + \frac{1}{\alpha\beta}. \quad (19)$$

The $M$-distribution is a generic statistical model which unifies several existing distributions used for atmospheric turbulence [130], [131]. Various distribution models like log-normal, $K$, Gamma-Gamma, negative exponential can be generated by using the unified $M$-distribution model. Although Gamma-Gamma distribution is most widely used to study the performance of FSO system, however, in a recent work proposed by Chatzidiamantis et al. [132], Double-Weibull distribution is suggested to be more accurate model for atmospheric turbulence than the Gamma-Gamma distribution, particularly for the cases of moderate and strong turbulence. Similar to the Gamma-Gamma model, this distribution is based on the theory of doubly stochastic scintillation and considers irradiance fluctuations as the product of small scale and large scale fluctuations which are both Weibull distributed. A very latest turbulence model proposed in [133] is Double Generalized Gamma (Double GG) distribution which is suitable for all regimes of turbulence and it covers almost all the existing statistical models of irradiance fluctuations as special cases. Another distribution proposed in [134], is Exponentiated Weibull (EW) distribution for modeling irradiance fluctuations in the presence of aperture averaging. Though Gamma-Gamma model is accepted to be valid in all turbulence regimes for a point receiver, however, it does not accurately fit the irradiance statistics when aperture averaging is used [135]. In [136], the EW distribution is proved to be valid with fully and partially coherent beams under all turbulence conditions in the presence of aperture averaging. Average channel capacity for all turbulence regimes over EW distributed non-Kolmogorov turbulent channel is presented in [137].

(III) **Beam divergence loss:** As the optical beam propagates through the atmosphere, a beam divergence is caused by diffraction near the receiver aperture. Some fraction of the transmitted beam will not be collected by the receiver and that will cause beam divergence loss/geometrical loss. This loss increases with the link length unless the size of the receiver collection aperture is increased or receiver diversity is employed. In general, an optical source with narrow beam divergence is preferable. But narrow beam divergence causes the link to fail if there is a slight misalignment between the transceivers. Therefore, an appropriate choice of beam divergence has to be made in order to eliminate the need for an active tracking and pointing systems and at the same time, reduce the beam divergence loss. The beam divergence is a very critical design parameter for ground-to-satellite FSO uplink. It has to be broad enough to reach the satellite with high degree of probability and at the same time ensure that the transmit power requirements are within range. The downlink beam divergence drives the attitude control requirements and determines the overall throughput of communication link. In [138], divergence and geometrical loss calculations are performed for analyzing the feasibility of FSO links between the Earth and satellites, the Earth and aircraft, aircraft and satellites, the Earth and moon, the Earth and Mars, and the Earth and the edge of the solar system. Fine steering and pointing control requirements are analyzed in [139] to tailor the specific requirement of beam divergence and FOV in case of FSO uplink and downlink.

(IV) **Background noise and sky radiance:** The main sources of background noise are: (a) diffused extended background noise from the atmosphere, (b) background noise from the Sun and other stellar (point) objects and (c) scattered light collected by the receiver [140]. The background noise can be controlled by limiting the receiver optical bandwidth. Single optical filter with very narrow bandwidth in the order of approx. 0.05 nm can be used to control the amount of background noise. Some of the design considerations while selecting narrow band optical filter are angle of arrival of the signal, Doppler shifted line width of the laser and various temporal modes. In addition, the other sources of noise in the FSO system are detector dark currents, signal shot noise and thermal noise. Total noise contribution is sum of background noise and noise due to other sources. Light from luminous bodies like the Sun, the moon or other fluorescent objects produces shot noise and it interferes with the noise present at the detector. Both the detector noise and ambient noise contribute to total receiver noise and this result in some noise flicker or disturbances. Further, large amount of scatterers and stray background noise may lead to damage or saturation of detectors and sensor. In order to minimize the noise from ambient light sources, the FSO system should be operated at higher wavelengths. The solar irradiance spectrum ranges from 300 nm to more than 2000 nm showing peak at around 500 nm. Thereafter, the pattern decreases with the increase in the wavelength [141]. It is to be mentioned that the sky radiance is present both during day-time and night-time whose magnitude is driven by the receiver's LOS Sun angle, the receiver's altitude, molecular and aerosol concentration and cloud density. The amount of sky radiance collected by the receiver is given as [85]

$$P_{sky} = S\left(\lambda, \theta, \varphi\right) \frac{\pi D_R^2 \Omega \Delta \lambda}{4} \qquad (20)$$

where $S$ is the sum of all the scattering sources in all the $N$ atmospheric layers, $D_R$ is the receiver diameter, $\Omega$ is FOV of receiver in steradians and $\Delta \lambda$ is the bandwidth of a narrow bandpass filter. The sky radiance is greatly dominated by the aerosol scatter within 30° angular distance of the receiving telescope from the Sun. With the increase of angular distance, molecular scattering becomes more significant. The amount of sky radiance greatly depends upon the phase angle defined by geometrical location of Sun, Earth and the receiver. For very small Sun-Probe-Earth (SPE) and Sun-Earth-Probe (SEP) angles, a sufficient amount of stray background rejection is essential to improve the system performance. In [142], sky radiance measurements were carried out at three different locations (Thessaloniki, Izana and Innsbruck) which shows the deviation in their measured values that varies between 3 and 35% depending upon wavelength, location and instrument. An atmospheric radiance model that takes into account clear sky radiance, radiance for a blackbody at the temperature of the near-surface air and angular dependance between telescope and Sun is studied in [143].

**Mis-alignment or pointing loss:** The optical beam used in the FSO communication is highly directional with very narrow beam divergence. Also, receivers used in FSO links have limited FOV. Therefore, in order to have 100% availability of FSO links, it is very essential to maintain a constant LOS connection between the transmitter and receiver. A slight misalignment can lead to failure of FSO link. It is very essential to maintain pointing and acquisition throughout the duration of communication. Pointing loss could arise due to many reasons such as satellite vibration or platform jitter or any kind of stress in electronic or mechanical devices [144]. The effect of satellite vibration in the FSO system is described in [145]–[151]. Pointing error can also be caused due to the atmospheric turbulence-induced beam wander effect which can displace the beam from its transmit path [152]. In any of the cases, pointing error will increase the chances of link failure or can significantly reduce the amount of received power at the receiver resulting in a high probability of error. In order to achieve a sub micro-radian pointing accuracy, proper care has to be taken to make the assembly vibration free and maintain sufficient bandwidth control and dynamic range in order to compensate for residual jitter [153].

The pointing error is described statistically by Gaussian distribution for both azimuth and elevation uncertainties. The radial pointing error angle (without bias) is the root sum square of the elevation $\theta_V$ and azimuth $\theta_H$ angles such that $\theta = \sqrt{\theta_H^2 + \theta_V^2}$. Assuming the standard deviation of elevation and azimuth angles are equal ($\sigma_\theta = \sigma_V = \sigma$) and further assuming that azimuth and elevation processes are zero-mean, independent, and identically distributed, then the radial pointing error angle can be modeled as a Rician density distribution function given by

$$f(\theta, \phi) = \frac{\theta}{\sigma^2} \exp\left(-\frac{\theta^2 + \phi^2}{2\sigma^2}\right) I_0\left(\frac{\theta\phi}{\sigma^2}\right), \quad (21)$$

where $I_0$ is the modified Bessel function of order zero and $\phi$ the bias error angle from the center. When $\phi$ is zero, the above equation leads to the well-known Rayleigh distribution function for pointing error angles and is given by [154]

$$f(\theta) = \frac{\theta}{\sigma^2} \exp\left(-\frac{\theta^2}{2\sigma^2}\right). \quad (22)$$

Total pointing error, $\sigma_p$ is sum of tracking error, $\sigma_{track}$ and point ahead error, $\sigma_{pa}$ i.e., $\sigma_p = \sigma_{track} + \sigma_{pa}$. Tracking error is primarily due to the noise associated with tracking sensors or due to disturbances arising from mechanical vibration of the satellite. Point-ahead-error occurs if the calculation of point-ahead-angle (PAA) did not allow sufficient transit time from satellite-to-ground and back again (more details on PAA is given later in this Section). It could be due to error in Ephemeris data or point ahead sensor error or calibration error or waveform deformation. Pointing error loss is more when tracking LEO than GEO satellite [155], [156]. Also, loss due pointing error is more significant at visible wavelength and decreases at higher wavelength due to an inherent broadening of beam. Pointing error has significant impact on BER performance of the FSO system. Fig. 4 shows the BER performance in the presence of random jitter. In [157], rms standard deviation of the pointing error is evaluated for satellite-borne FSO communication system. A model is proposed in [158], to compensate for the tracking and pointing losses due to high amplitude vibrations between space-borne satellites. Beam wander-induced pointing error for different intensity modulation schemes for ground-to-satellite FSO links is presented in [159]. More literature on mis-alignment or pointing error in FSO links due to atmospheric turbulence can be found in [160]–[162].

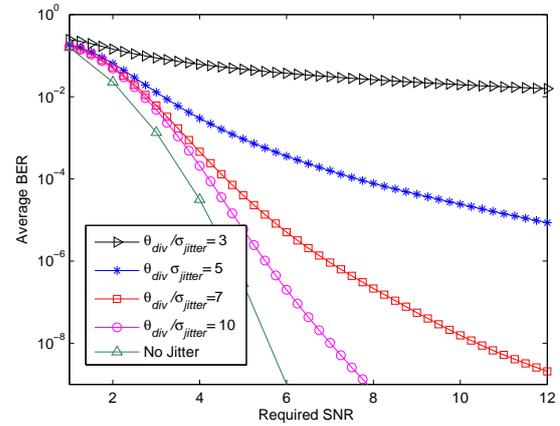

Figure 4. BER vs SNR for different values of ratio of beam divergence angle to random jitter [120]

(V) (VI) **Cloud blockage:** The presence of opaque clouds may occasionally disrupt the signal or completely block the optical signal from ground-to-satellite or satellite-to-ground rendering the LOS communication useless. These intermittent blockages can last from few seconds to several hours depending on the geographical location and season. They can lead to wandering off the downlink signal from the desired position if the space-borne system relies on uplink beacon signal for tracking and pointing. Cumulus clouds may appear alone or in clusters in the altitude range from 200 to 6500 feets above the ground and hinders the space-based optical communication [163]. Clouds offer significant attenuation as high as tens of dB and therefore require necessary actions to combat the signal loss due to cloud coverage. Implementation of spatial and temporal diversity can minimize the amount of attenuation due to clouds. International satellite cloud climatology project (ISCCP) provides database for weather conditions that can be utilized for the optical site section of various ground stations [164]. National climatic data center in Asheville, North Carolina provides information on cloud cover, visibility and other parameters over 1000 sites in United States [165]. This database is useful for the site selection to develop optical space network (OSN) and serves as a guide to build various models of real time dynamic cloud behaviors. These models are used to provide cloud-free line of sight (CFLOS) [166] or cloud-free arc (CFARC) probabilities for any site. In order to provide CFLOS probability, multiple ground stations are placed at a low cloud coverage area with a separation of at least 350 km to ensure uncorrelated CFLOS [167]. These multiple ground stations have a communication infrastructure to support relay-satellite-communication throughout the world. Temporal characteristics of optical beam propagation through the clouds using Monte Carlo simulations are studied in [168]. Mathematical modeling of the FSO communication system over cloudy channel based on adaptive optical transmitter and receiver is discussed in [169]. Infrared cloud imaging to determine day-night cloud detection, cloud optical depth and optical attenuation in cloud channel is discussed in [170].

(VII) **Atmospheric seeing:** The perturbations of the optical beam associated with coherence length of the atmosphere, $r_0$ is referred as atmospheric seeing effect. When $r_0$ is significantly smaller than the receiver aperture diameter $D_R$, then it leads to the blurring of the received signal which is known as astronomical seeing which is given as $\lambda/r_0$ [171]. For a perfect optical collection system, the spot size of the received signal in the focal plane of the receiver is expressed as $(2.44F\lambda/D_R)$ where $F$ is the focal length of receiver collecting optics. When the optical beam propagates through atmosphere, then $D_R$ is replaced by $r_0$ and therefore, the related signal spot size at the focal plane is increased by the ratio $D_R/r_0$ which effectively leads to an increase in the background noise. Also, larger FOV at the receiver can limit the electrical bandwidth of the receiver, thereby limiting the data rate. This problem can be taken care of by use of adaptive optics or array detectors.

(VIII) **Angle-of-arrival fluctuations:** Due to the presence of turbulence in the atmosphere, the laser beam wavefront arriving at the receiver will be distorted. This will lead to spot motion or image dancing at the focal plane of the receiver. This effect is called angle-of-arrival fluctuation. However, this effect can be compensated by the use of adaptive optics or fast beam steering mirror. The variance of angle-of-arrival fluctuations, $\langle\beta^2\rangle$ is expressed as [85]

$$\langle\beta^2\rangle = 2.91\left[\int_{h_o}^{H}C_n^2(h)\,dh\right]D_R^{-1/3}\sec(\theta), \quad (23)$$

which is further simplified to yield

$$\langle\beta^2\rangle = 0.182\left(\frac{D_R}{r_0}\right)^{5/3}\left(\frac{\lambda}{D_R}\right)^2, \quad (24)$$

where $D_R$ is the diameter of collecting lens and $r_0$ is the coherence length of the atmosphere.

Besides the above mentioned factors, there could be other reasons for link failure. Since FSO system requires LOS communication, any kind of physical obstruction can block the beam path and cause short and temporary interruptions of the received signal. This adverse effect can be taken care of by proper choice of system design parameters like beam divergence, transmitter power, operating wavelength, transmitter and receiver FOV.

*B. Inter-satellite Links*

Inter-satellite FSO links are not subject to weather conditions or cloud outages as the satellite orbits are far above the atmosphere. In this case, the major challenge is caused by the acquisition and tracking as the two satellite move with different relative velocity. As inter-satellite or inter-orbital links have to cover larger distances, therefore the transmission scheme has to be power-efficient with good sensitivity at the receiver. Therefore, phase-coherent techniques like homodyne or heterodyne are preferred over direct detection methods in inter-satellite FSO links. These techniques provide very good receiver sensitivity and deliver high capacity links. A homodyne BPSK transmission between LEO-LEO with 5.6 Gbps is the highest rate published till date in [172], [173]. The European space agency has developed European Data Relay Satellite System (EDRS) that successfully demonstrated 1.8 Gbps link between Alphasat in GEO and Sentinel-1 in LEO in late 2014 [174], [175]. A theoretical analysis of the inter-satellite FSO link is performed for 1000 km distance at 2.5 Gbps in [15]. Although, space FSO links are not subject to atmospheric and weather limitations, however, they are limited by other challenges like PAA, doppler shift, acquisition and tracking, background radiations and satellite platform stability. These challenges are discussed as follows:

(I) **Point-ahead-angle**: Due to the relative motion between the transmit and receive terminals, the return signal is required to be an offset from the apparent beacon location so that it effectively hits the receiver at a proper spatial temporal location. This pointing offset is called PAA. It depends upon the relative velocity between the two satellites and is used to compensate for the travel time over the long cross link distances. For deep space optical links, PAA is of the order of hundreds of micro-radians and for inter-satellite/ground-to-satellite links, its value is typically tens of micro-radian. Fig. 5 illustrates the concept of PAA where the optical beam sent by LEO at time "$T$" will receive a return beam from GEO at time "$T + \triangle T$". In order to provide a more precise pointing accuracy, the transmit and receive terminals are frequently aligned using link maintenance control and tracking algorithm [176], [177]. Generally, an effect called point ahead angular anisoplanatism is observed if the PAA is much larger than the isoplanatic angle from the tracking direction. It is generally caused by a mismatch of the laser beam from the beacon path. The mismatch is characterized by the angular pointing error and the turbulence is characterized by the isoplanatic angle given as $\theta_0^{-5/3} = 2.91k^2\int_0^L h^{5/3}C_n^2(h)\,dh$ [178]. Detailed study of anisoplanatism effects in space FSO communications can be found in [179], [180].

(II) **Doppler Shift:** The change in the frequency of the received signal due to the relative motion between the source and the receiver leads to Doppler effect. It happens in inter-orbit satellite links where the satellite in lower orbits travel faster than at higher orbits. The amount of shift that has to be compensated in data relay systems is approximately $\pm 7.5$ GHz between LEO and GEO transmissions. The shift can be even more for two LEOs propagating in opposite directions. Doppler shift results in approximately 140 kHz frequency shift in a 2 GHz clock used for signaling [70], [181]. This may require a wide frequency tuning range for optical input filters or local oscillator (LO) lasers in case of coherent optical link [155]. This Doppler shift has to be taken into consideration for reliable FSO inter-satellite links. The Doppler shift reaches a maximum value when the radial component of the relative velocity reaches its maximum. Negligence of Doppler effect results in loss of data and frequency synchronization issues at the receiver end. A conventional solution to reduce Doppler shift can be the integration of an optical phase-lock loop (OPLL) technique with the LO laser [182]. In this case, cooperative frequency tuning between both the transmitter and LO at the receiver is implemented to combat the effect of Doppler shift. The frequency acquisition process begins with the LO being tuned with an open frequency control loop towards the transmitter until a beat signal (difference of transmitter and LO frequency) is obtained. Once the beat signal is obtained, frequency tracking mode initiates to lock on to the heterodyne frequency control loop or homodyne loop. Another solution is the use of the optical injection locking (OIL) technique, however, there is a drawback that the locking frequency range is limited to typically 1 GHz or less [183]. A combination of OPLL technique with an OIL technique i.e., use of an optical injection phase-lock loop (OIPLL) technique, which is also used to lock the frequency and phase under Doppler shift conditions. A pilot-carrier coherent LEO-to-ground downlink using OIPLL Doppler frequency shift is demonstrated in [184].The effect of Doppler shift in homodyne inter-satellite communication link is analyzed in [185]. Optimum bandwidth of the optical filter is determined for the entire range of frequency shifts to improve the system performance. The frequency shift induced by Doppler effect is characterized for LEO satellite constellations with optical inter-satellite links in [186]. Doppler shift frequency compensator for coherent homodyne receiver with optical phase lock loop is investigated in [187].

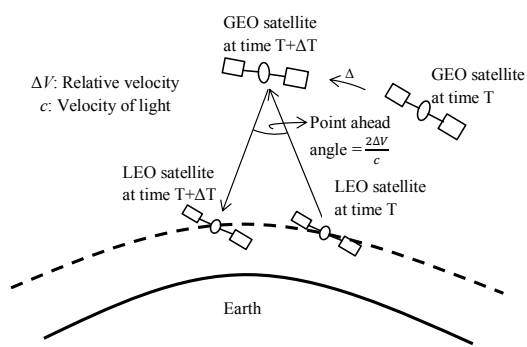

Figure 5. Concept of PAA in space communication

(III) **Satellite vibration and tracking:** Acquiring and tracking the received signal LOS through disturbances that are generated either by the satellite or by the on-board laser communication sub-assembly is a challenging task for inter-satellite FSO links. Various externally generated disturbances by the satellite are due to solar panels, momentum wheels, gimbal packages, thrusters, etc. Various noise sources generated by on-board laser communication systems are Relative Intensity Noise (RIN), thermal noise, dark current shot noise, signal shot noise, and background shot noise. These satellite vibrations plus noises cause deviation in the direction of the transmitted beam and result in misalignment between two satellites. A good understanding of internal and external disturbances improves the satellite structure modeling and pointing budget requirements for inter-satellite FSO links. The direction of the laser beam is corrected with the help of beam steering mirrors. Broadly, tracking systems are categorized into two categories: (i) the system that derives the track information from communication signal and (ii) system that uses a seperate beacon signal for tracking. Various tracking techniques such as dc tracking, pulse tracking, square law tracking, coherent tracking, tone tracking, feed-forward tracking and gimbal tracking are implemented for inter-satellite FSO links. Coherent tracking uses high front end LO gain to compensate for large background noises that limit the receiver sensitivity [188], [189]. Gimbal tracking is generally used where large beam divergences are available as the gimbal jitter is large enough to support narrow beams. Feed-forward tracking uses only one beam steerer in the transmit path and requires large linear range on the receive detector. This requires the tracking system to have large received spot size that will eventually reduce the sensitivity of the receiver. A decrease in sensitivity is compensated by an increase in transmit power. A feed-forward vibration compensation model is compared with feedback model for the inter-satellite FSO link in [190], [191]. In [192], a model is proposed that adapts the bandwidth and transmitter gain to compensate for the satellite vibration. In another work presented in [193], the gain of the telescope is varied adaptively using phased array techniques to maximize the communication performance for new vibration level. More literature about compensation of satellite vibration and tracking issues can be studied in [194]–[196].

(IV) **Background noise sources:** The noise sources for inter-satellite link depends on the detection technique and if the system is optically pre-amplified or not. For direct detection receiver, major noise contribution is often from detector (bulk dark current), receiver amplifier (pre-amplifier noise and thermal noise) and shot noise due to the signal itself. For coherent detection, local oscillator shot noise dominated all the other noise sources present in the system. Other background noise sources consist of stellar and celestial radiant fluxes. It also consists of scattered light from optics structure falling on the detector. The background power collected by the receiver due to diffused, stellar and scattered noise is given by

$$P_B = \begin{cases} H_B \Omega_{FOV} L_R A_R \Delta\lambda_{filter} & \text{(Diffused source)} \\ N_B L_R A_R \Delta\lambda_{filter} & \text{(Stellar or point source)} \\ \gamma I_\lambda \Omega_{FOV} L_R A_R \Delta\lambda_{filter} & \text{(Scattered noise source)} \end{cases}$$
(25)

In the above equation, $H_B$ and $N_B$ are the background radiance and irradiance energy densities of large diffused angular sources and point sources, respectively. The term $H_B$ is expressed in units of W/m$^2$/sr/Å and $N_B$ in terms of W/m$^2$/Å. The parameter $\Omega_{FOV}$ is the solid angle receiver FOV, $L_R$ is the transmission loss of the receiver optics, $A_R$ is the effective area of the receiver and $\Delta\lambda_{filter}$ is the bandwidth of optical BPF in the receiver. In case of scattered noise source, $\gamma$ represents the atmospheric attenuation coefficient and $I_\lambda$ is the exo-atmospheric (region of space outside the earth's atmosphere) solar constant (0.074 W/cm$^2\mu m$). A strong background source near the receiver FOV can lead to significant scattering. Degradations caused by radiating celestial bodies other than the Sun are generally negligible except when receiving optics are directly pointing the Sun. The major source of background noise is due to scattering when an optical receiver design has its optics under direct exposure to sunlight.

## III. Acquisition, Tracking and Pointing

In case of ground-to-satellite, inter-satellite and satellite-to-ground FSO communication links, pointing and focusing of optical beam become a critical issue. This is due to narrow beam divergence of the optical beam and platform jitter disturbances [197]. The performance of the space optical link is limited due to vibration of the pointing system caused by two stochastic mechanisms: (i) tracking noise created by the electro-optic tracker and (ii) vibrations created by internal satellite mechanical mechanisms. The effect of satellite vibration and platform jitter is described in [198]–[201]. Misalignment due to any of these reasons may result in link failure or severely degrade the performance of the system. This section will cover various issues involving acquisition and tracking of narrow optical beam in order to achieve a stable LOS communication between the two terminals.

The first step to establish a space optical link is acquisition process which involves the transmitter scanning with its narrow beacon signal over an uncertainty area. The beacon signal should have a sufficient peak power and low pulse rate to help the receiver locating the beam in the presence of large background radiations [202]. The coarse detection is done during the acquisition mode with large FOV. This beacon signal is detected by the position sensitive detector at the receiver which is also searching simultaneously over its FOV for the beacon signal as shown in Fig. 6.

Once the beacon signal is detected, the receiving terminal makes use of beam steering elements to point a steady beacon signal towards the initiating terminal often offset by fixed PAA. Fig.7 shows the basic concept of ATP for a space FSO communication system. The uplink signal from the ground

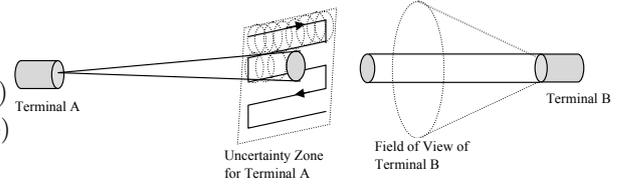

Figure 6. Stare/Scan acquisition technique where one terminal (Terminal A) slowly scans its uncertainty region while other terminal (Terminal B) stares through over its FOV

station is collected with the help of telescope optics and given to the dichroic beam splitter 1. This beam splitter will reflect all the incoming signal to the beam splitter 2 that will further direct the signal to the communication detector or ATP subsystem in accordance with the wavelength of the incoming signal. In case of beacon signal, the beam splitter 3 will help in focussing the image of the ground station to a point on the focal pixel array (FPA). The location of this point on the array represents the direction from the received beacon signal relative to the telescope's axis (the center of the array). The size of the FPA determines the FOV of the telescope and is large enough to accommodate the initial pointing uncertainties. Up till this point, the first portion of the spatial acquisition process is complete wherein the position of ground station relative to the space craft is captured on the FPA. Next step is to direct the beam from an on-board mounted laser satellite to the ground-based receiver by forming a LOS link.

After the beacon signal is received, the controller logic on the satellite begins the process of narrowing its FOV until both systems have locked on to each other's signal. For this reason, there must be adequate signal level for initial detection at the distant space-borne satellite and sufficient received signal energy to allow closed loop tracking to begin. Typically, the transition from acquisition to tracking phase takes less than one second. First, the coarse tracking is performed by means of a control loop driving the coarse pointing mirror. Later, the controller logic on the satellite commands the fine beam steering element to keep the received signal bore sighted on the detector [203], [204]. This angular control of beam steering mirror is achieved by the error signal that is the difference between the current position of the on-board mounted laser satellite and the beacon signal from the ground station. The error signal will keep on driving the fine beam steering mirror till it reaches to a minimum value which implies that both the on-board and ground station lasers are now aligned to each other. The control loop tracking bandwidth should be ≥1KHz to accommodate for satellite vibrations [205]. Charge-coupled device (CCD) arrays, complementary metal oxide semiconductor (CMOS) arrays, or quadrant avalanche photodiodes (QAPDs) can be employed as tracking sensors. CCDs have wider FOV for coarse tracking whereas QAPDs are much faster and accurate for fine tracking [205].

The transmitting and receiving terminals use steady beacon to cooperatively increase the optical signal between them in order to increase the track loop bandwidth for improved pointing accuracy. Therefore, the choice of beam divergence is very critical to provide enough received signal energy to support the initial detection of the target within the required acquisition time and at the same time allows the transition to narrow beam tracking. Lot of work has been carried out for optimization of acquisition time in case of inter-satellite/ground-to-satellite optical link [206]–[211] as it determines the performance of the acquisition system. In [212], a wavelength diversified FSO link consisting of three wavelengths i.e., 1.55 $\mu$m, 0.85 $\mu$m and 10 $\mu$m has been modeled from ground to mobile aerial platform communication. Based on the transmission properties, a method for minimizing link acquisition times through the exploitation of various properties of each wavelength was analyzed.

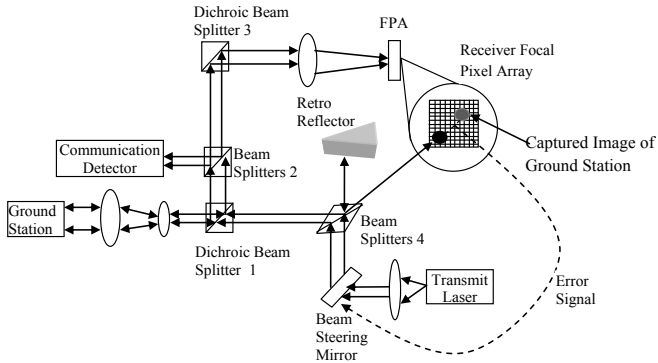

Figure 7. Block diagram of the ATP system between ground station and on-board satellite [85]

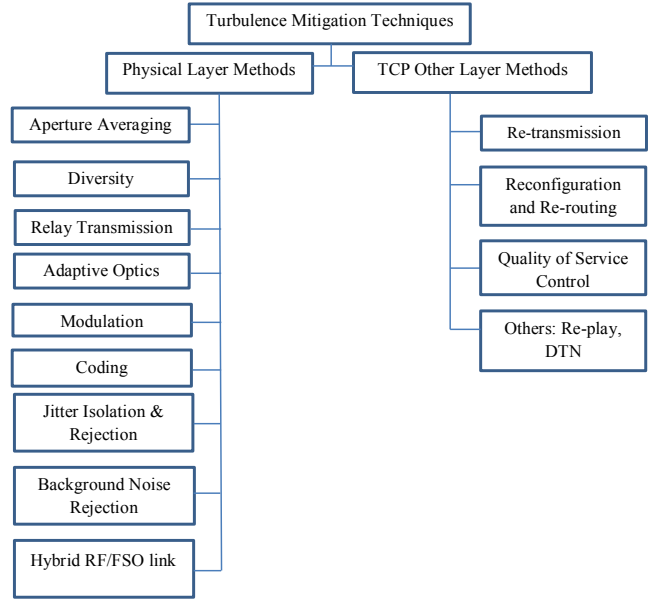

Figure 8. Various techniques for mitigating atmospheric turbulence

In case of a space optical communication, pointing accuracy is accomplished using an hybrid pointing architecture that makes use of inertial sensors, celestial reference and an uplink beacon. In some cases, beaconless pointing using celestial reference source only [213], [214] is used to provide a desired pointing reference. This architecture provides a simplified operation with considerable power saving as it eliminates the need of extra beacon signal.

## IV. MITIGATION TECHNIQUES

The atmospheric channel causes degradation in the quality of the received signal which deteriorate the BER performance of the FSO system. In order to improve the reliability of the FSO system for all weather conditions, various types of mitigation techniques are employed. Mitigation techniques can be used either at physical layer or at other TCP layers (network or transport layers). Multiple beam transmissions, increasing receiver FOV, adaptive optics, relay transmission, hybrid RF/FSO etc., are some of the mitigation techniques used at physical layer. Packet re-transmission, network re-routing, quality of service (QoS) control, data re-play are some of the methods used in other TCP layers in order to improve the performance and availability of FSO systems. Fig. 8 gives various mitigation techniques used in space-based optical communication.

### A. Physical Layer Methods

(I) **Aperture Averaging:** This technique is used to mitigate the effect of atmospheric turbulence by increasing the size of the receiver aperture that averages out relatively fast fluctuations caused by the small-size eddies and helps in reducing channel fading. The parameter that quantify reduction in fading due to aperture averaging is called aperture averaging factor, $A$. The parameter $A$ is defined as the ratio of variance of the signal fluctuations from a receiver with aperture diameter $D_R$ to that from a receiver with an infinite small aperture i.e.,

$$A = \frac{\sigma_I^2(D_R)}{\sigma_I^2(0)}. \quad (26)$$

In case of a satellite-to-ground FSO link, aperture averaging factor for both plane and spherical waves in weak turbulence is approximated as [215]

$$A = \left[1 + 1.1\left(\frac{D_R^2}{\lambda h_0 \cos\theta}\right)^{7/6}\right]^{-1} \quad (27)$$

where $h_0$ is the scale height of the turbulence and other parameters are defined earlier. Therefore, increasing the aperture diameter reduces atmospheric scintillation and improves the BER performance of the system. For the space-based optical link, the plane wave propagation model is mostly appropriate for satellite-to-ground transmissions, whereas the spherical propagation model is suitable for ground-to-satellite transmission links. . Various literature on aperture averaging is found in [215]–[221]. In case of an aperture diameter larger than the coherence length of the atmosphere, the irradiance statistics appear to be log-normal. Besides widely accepted log-normal and Gamma-Gamma distributions, a new distribution family, namely Exponentiated Weibull (EW) distribution is proposed in [222], which offers an excellent fit for simulated and experimental data under weak and moderate atmospheric turbulence conditions.

For space-based optical links, various experimental and theoretical studies have been carried out to analyze the effect of the detector size on atmospheric turbulence. In [223], the performance of the optical satellite communication link for different modulation schemes using aperture averaging has been investigated. It was found that an improvement in the system performance employing On Off Keying (OOK) modulation is higher as compared to subcarrier binary phase shift keying (SC-BPSK) and subcarrier quadrature phase shift keying (SC-QPSK) modulation schemes. BER performance for coherent (sub-carrier BPSK) and non-coherent (OOK) modulation schemes using aperture averaging for the ground-to-satellite FSO communication is investigated in [224]. In [215], performance improvement for the satellite-to-ground link has been observed by using aperture averaging. The synergy of the channel coding and aperture averaging is analyzed for downlink optical signal from LEO satellite [225]. It was observed that the aperture averaging shows good results for values of zenith angles. A comparison of terrestrial and satellite FSO links using various coherent modulation schemes in [226] suggested that the performance improvement due to aperture averaging is only effective in terrestrial communication and satellite downlink. The advantage of aperture averaging for the uplink satellite communication is not available as the phase coherence radius is much larger than receiving aperture on the satellite.

It should be noted that an increase in the receiver aperture area will also increase the amount of background noise collected by the receiver. Therefore, an optimum choice of aperture diameter has to be made in order to increase the power efficiency in the FSO system.

(II) **Diversity:** Diversity technique for mitigating the effect of turbulence in the atmosphere can operate on time, frequency and space. In this case, instead of a single large aperture, an array of smaller receiver aperture is used so that multiple copies of the signal that are mutually uncorrelated can be transmitted either in time or frequency or space. This will improve the link availability and BER performance of the system. It also limits the need of active tracking due to laser misalignment. GOLD demonstration in 1998 showed that the scintillation index was drastically improved with four 514.5 nm multiple beams for uplink transmission. It was reported that the value of scintillation was 0.12 with two beams, however, its value reduces to 0.045 with four beams [227]. Reduction in the intensity variance by a factor of square root of the number of transmitting antennae is observed in [228], [229] for ground-to-satellite FSO uplink. In order to achieve the full advantage due to spatial diversity, the antennae separation at transmitter or receiver should be at least or greater than the coherence length of the atmosphere to make the multiple beams independent or at least uncorrelated. In [230], the atmospheric scintillation for the ground-to-satellite FSO link is analyzed using multiple uplink Gaussian beams. An average spatial correlation is derived as a function of beam waist, beam separation and beam wander variance. Also, the gain due to diversity is more pronounced at a high turbulence level than at lower values [231], [232].

In case of receiver diversity (SIMO- single input multiple output), diversity gain is achieved by averaging over multiple independent signal paths. The signals can be combined at the receiver using selection combining (SC) or equal gain combining (EGC) or maximal ratio combining (MRC). SC is simpler as compared to other two, but gain in this case is low. The gain achieved through MRC is slightly higher than EGC, but at the expense of complexity and cost. Therefore, implementation of EGC is preferred over MRC due to its simplicity and comparable performance [233], [234]. For optical downlink from deep space, multiple receivers are considered in [235] which are placed in an orbit above the atmosphere. However, it becomes a costly affair to launch and maintain orbiting space receiver as compare to ground-based receivers. For transmit diversity (MISO- multiple input single output), special space time codes such as optical Alamouti code is used [128], [236]. This code is designed for only two transmit antennae but can be extended to more number of antennae. FSO MIMO (multiple input multiple output) system performs well if the beams are independent or uncorrelated. Otherwise, the performance of FSO system is going to degrade. The performance of optical MIMO and RF MIMO systems are almost equivalent. It increases the channel capacity of the system almost linearly with the number of transmitting antenna. In the case of weak atmospheric turbulence, the outage probability of Gaussian FSO channel is proportional to $[\log(\text{SNR})]^2$ term whereas for moderate to strong turbulence, it is proportional to $[\log(\text{SNR})]$ [237]. The concept of virtual MIMO (V-MIMO) is applied to the constellation of multiple HAPs for providing high capacity broadband links in the presence of cloud blockage or turbulent atmosphere. Time diversity with or without codes has also been proven to mitigate channel fading in the FSO system. This type of diversity is applicable to time selective fading channels which allows repetitive symbols to be transmitted over different coherence time period. If data frame length exceeds the channel coherence time, then diversity can be employed by either coding or interleaving. It is observed that in the presence of time diversity, convolutional codes are a good choice for weak atmospheric turbulence and Turbo-codes provide a significant coding gain for strong turbulence conditions [238]. The combination of encoding abilities of punctured digital video broadcast satellite standard (DVB-S2) and LDPC is used to mitigate the atmospheric turbulence by using time diversity in the FSO system [239].

(III) **Relay Transmission:** Relay-aided transmission is an effective technique for combating the effect of

turbulence in the FSO communication [240]–[243]. It is a form of distributed spatial diversity that enables multiple terminals to share their resources by a cooperative communication so that a virtual antenna array can be built in a distributed fashion. Here, instead of using multiple apertures at the transmitter or receiver end, a single antenna is capable of achieving a huge diversity gain. Relay transmission provides good a performance advantage when operated at a high SNR such that the received signal strength is sufficiently high than the signal dependent shot noise and atmospheric fading. Otherwise, in case of a low SNR, relays will be forwarding just the noisy replicas of the information they received. The outage probability and ergodic capacity for relay transmission in the FSO system provide significant improvement over direct transmission in turbulent atmospheric channel. The performance of hybrid satellite-terrestrial FSO link is analyzed using amplify and forward relaying in [243]. The average symbol error rate and diversity order is obtained using M-ary phase shift keying. Further, the work is extended in [244], where the system performance was analyzed in the presence of co-channel interference. The optical downlink from LEO/GEO satellite-to-ground station is completed via HAP relay link. The network of interconnected HAP provide almost full availability in all weather conditions. Recently, the performance analysis of the relay-aided multi-hop FSO system over EW fading channels with pointing error has been analyzed in [245]. The placement of relays plays a vital role in achieving a high order diversity gain using the cooperative FSO communication and is studied in [246]–[249]. Further, in order to take a full capacity advantage of the optical carrier, all optical relaying FSO systems are studied in various literatures [250]–[252]. The effect of noise caused by the background light and amplifier in all-optical FSO systems is reduced by using optical regenerate-and-forward (ORF) relaying technique [251] or optical amplify-and-forward (OAF) relaying technique [252]–[256].

(IV) **Adaptive Optics:** Adaptive optics (AO) is used to mitigate the effect of atmospheric turbulence and helps to deliver an undistorted beam through the atmosphere. AO system is basically a closed loop control where the beam is pre-corrected by putting the conjugate of the atmospheric turbulence before transmitting it into the atmosphere [257]–[259]. An increase in transmit power or using diversity can improve the performance of the FSO system. But in order to have further improvements in SNR with a reduced transmit power requirement, AO have proved to be very beneficial. The implementation of the AO system in Compensated Earth-Moon-Earth Retro-Reflector Laser Link (CEMERLL) showed significant improvement in the received SNR [28]. The AO system makes use of wavefront sensor, wavefront corrector and deformable mirrors either at the transmitter or at the receiver optics to compensate for the phase front fluctuations. Here, a part of the received signal is sent to the wavefront sensor that produces a control signal for the actuators of the wavefront corrector as shown in Fig. 9. However, a real time wavefront control using conventional AO approach becomes quite difficult for very strong turbulent conditions [28], [259]–[261]. In such situations, a non-conventional AO approach is used which is based on the optimization of the received SNR or any other system's performance metric [262], [263]. Earlier, this non-conventional approach was largely disregarded as it imposed serious limitations for the control bandwidth. But later with the development of high bandwidth wavefront phase controllers e.g., deformable mirrors based on micro-electromechanical systems (MEMS) and with the development of new efficient algorithms, this approach is gaining popularity these days. The AO using MEMS for long range FSO systems is found in [264]–[267]. The use of adaptive optics for high capacity returns from LEO or GEO satellite-to-ground receiver is investigated in [268]. The tip/tilt tracking of the incoming wavefront amounts to the tracking of the spot focused on the detector. The rms turbulence-induced wavefront tip/tilt is independent of wavelength, increases with zenith angle and is proportional to $D_R^{-1/6}$ [269]. Tip/tilt correction is achieved using natural or artificial guide stars or a down-directed laser beam near an orbiting relay mirror provides as a reference for measuring or correcting the wavefront distortion [270]. A Strehl ratio model for tip/tilt and adaptive optics correction is presented in [271]. Designing of the AO system requires that its closed loop frequency should be at least four times Greenwood frequency [272] (in Hz) given by

$$f_G = \left[ 0.102 k^2 \sec(\theta) \int_{h_o}^{h_o+L} C_n^2 \cdot v_T(h)^{5/3} dh \right]^{3/5}, \quad (28)$$

where $v_T(h)$ is the traversal component of wind speed. This frequency tells the speed of AO system to respond to fluctuations due to atmospheric turbulence.

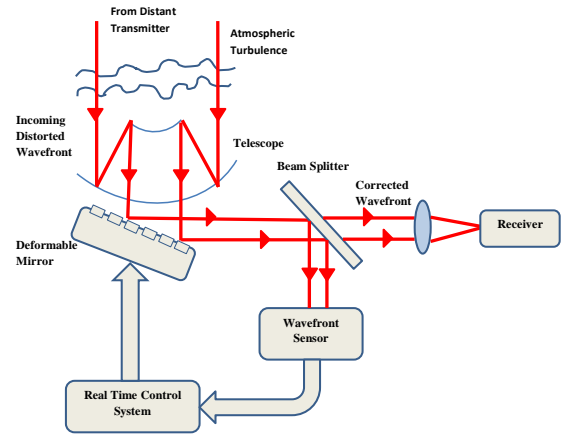

Figure 9. Conventional adaptive optics system

(V) **Modulation:** In the FSO communication, the choice of modulation schemes depends on two main criteria: optical power efficiency and bandwidth efficiency. Optical power efficiency can be measured by computing optical power gain over OOK provided both the modulation schemes have same euclidean distance, $d_{min}$. Power efficient modulation schemes are simpler to implement and are quite effective in mitigating the effect of the turbulence for low data rates. Bandwidth efficiency on the other hand, determines maximum data for a given link length with a particular modulation scheme.

In general, the FSO communication supports a variety of binary and multilevel modulation formats. Out of these two formats, binary level format is most commonly used due to its simplicity and high power efficiency. Most well known binary modulation schemes are OOK and PPM. OOK modulation scheme requires adaptive threshold in turbulent atmospheric conditions for best results [273]. Due to its simplicity, OOK modulation scheme is very popular in FSO communication systems and most commonly is deployed with IM/DD transmission and receive mechanism. The choice of duty cycle for intensity modulated signals will impact system design parameters such as transmission bit rate and channel spacing. The ability to maintain the transmission efficiency over a wide range of duty cycles, requires efficient pulse shaping at the transmitter. It will not only improve the sensitivity of the receiver but will also provide flexible multi-rate capabilities with simplified receiver design options [42]. In some cases, laser may require some recovery time after transmission of the pulse and this may impose a minimum delay between two pulses. Another detection technique used in OOK modulation scheme is maximum likelihood (ML) detection with perfect channel state information (CSI) [274]. However, due to its implementation's complexity, this detection technique didn't gain much popularity. Maximum-likelihood sequence detection (MLSD) can be employed when the receiver is having the knowledge of a joint temporal distribution of intensity fluctuations. Other detection techniques [275]–[277] used at the receiver are symbol-by-symbol maximum likelihood detection, blind detection, vertical Bell laboratories layered space time (V-BLAST) detection, etc.

In case of *M*-PPM, each symbol interval is divided into *M* time slots and a non-zero optical pulse is placed at these time slots while other slots are kept vacant. For long distance or deep space communications, *M*-PPM scheme is widely used because it provides a high peak-to-average power ratio (PAPR) that improves its average-power efficiency [85], [278]. Also, unlike OOK, *M*-PPM does not require an adaptive threshold. However, *M*-PPM scheme has a poor bandwidth efficiency at higher values of *M* and therefore, for bandwidth limited systems, multi-level modulation schemes are preferred. Here, the transmitted data can take multiple amplitude levels and most commonly used multi-level intensity modulation schemes are pulse amplitude modulation (PAM) [279] and quadrature amplitude modulation (QAM) [280]. However, the price paid for bandwidth efficiency is the reduction in power level. Therefore, these modulation schemes are not a good choice for high turbulent atmospheric conditions or power limited systems. It is reported in many literatures that in case of high background noise, *M*-PPM is considered to be an optimum modulation scheme on Poisson counting channel [281]–[283]. With the increase in order of *M* in *M*-PPM, the robustness against background radiations increases even further due to its low duty cycle and lesser integration interval of photodiode.

Owing to various advantages of PPM in the FSO communication, various variants of PPM have been developed aiming to enhance the spectral efficiency of the system. A natural extension of single pulse PPM is to use two or more pulses in each channel symbol to convey information and is termed as multi-pulse PPM [284]. This is accomplished by placing multitude (*K*) of pulses per symbol intervals in all possible ways among *M* slots and therefore, provides signal constellation whose size varies as $M^K$ (for large *M*) rather than linearly with *M*, as in the case of the conventional PPM. Other variants of PPM are Differential PPM (DPPM) [285], [286], Differential Amplitude PPM (DAPPM) [287], Pulse Interval Modulation (PIM) [288], Dual Header Pulse Interval Modulation (DHPIM) and Overlapping PPM (OPPM) [289], [290]. All these modulation schemes are obtained by simple modification to PPM to achieve improved power and bandwidth efficiency. In DPPM, the empty slots following the pulse in PPM symbol are removed, therefore, it reduces the average symbol length and improves the bandwidth efficiency. DPPM exhibits inherent symbol synchronization as every symbol ends with a pulse. However, for a long sequence of zeros there could be a problem of slot synchronization which is taken care of by making use of guard band immediately after the pulse has been removed. The performance of DPPM and PPM has been compared in terms of BER and power spectral density in [291], [292]. It was found that DPPM results in improved bandwidth as well as power efficiency than PPM for a fixed-average bit rate and a fixed available bandwidth.

On the other hand, DAPPM is a combination of DPPM and PAM. It is therefore, a multi-level modulation scheme where the symbol length varies from 1, 2,. . . , *M* and the pulse amplitude is selected from 1, 2,. . . , *A* where *A* and *M* are integers. Performance analysis of ground-to-satellite FSO system with DAPPM in a weak atmospheric turbulence is studied in [287]. PIM is an anisochronous (no fixed symbol structure) PPM technique in which each block of $\log_2 M$ data bits are mapped to one of *M* possible symbols. The symbol length is variable and is determined by the information content of the symbol. Every symbol begins with a pulse, followed by a series of empty slots, the number of which is dependent on the decimal value of the block of data bits being encoded. Mapping between source and transmitted bits of 4-PPM and 4-PIM is shown in Table VI. It requires only a chip synchronization and does not require symbol synchronization since each symbol is initiated with a pulse. It has higher transmission capacity as it eliminates unused time chips within each symbol.

| Source Bits | 4-PPM | 4-PIM |
|---|---|---|
| 00 | 1000 | 1(0) |
| 01 | 0100 | 1(0)0 |
| 10 | 0010 | 1(0)00 |
| 11 | 0001 | 1(0)000 |

Table VI
MAPPING BETWEEN 4-PPM AND 4-DPIM CHIPS

| Modulation Schemes | $M$-PPM | DPPM | DAPPM | DPIM | $DHPIM_a$ |
|---|---|---|---|---|---|
| Bandwidth (Hz) | $\dfrac{MR_b}{\log_2 M}$ | $\dfrac{(M+1)R_b}{2\log_2 M}$ | $\dfrac{(M+A)R_b}{2\log_2(MA)}$ | $\dfrac{(M+3)R_b}{2\log_2 M}$ | $\dfrac{(2^{\log_2 M-1}+2\alpha+1)R_b}{2\log_2 M}$ |
| PAPR | $M$ | $\dfrac{M+1}{2}$ | $\dfrac{M+A}{A+1}$ | $\dfrac{M+1}{2}$ | $\dfrac{2(2^{\log_2 M-1}+2\alpha+1)}{3\alpha}$ |
| Capacity | $\log_2 M$ | $\dfrac{2M\log_2 M}{M+1}$ | $\dfrac{2M\log_2(M\cdot A)}{M+A}$ | $\dfrac{2\log_2 M}{M+3}$ | $\dfrac{2M\log_2 M}{2^{\log_2 M-1}+2\alpha+1}$ |

Table VII
COMPARISON OF VARIANTS OF PPM MODULATION SCHEME
($A$: PULSE AMPLITUDE, $R_b$: DATA RATE AND $\alpha$: INTEGER)

In DHPIM, a symbol can have one of the two pre-defined headers depending on the input information as shown in Fig. 10. The $n^{th}$ symbol $S_n(h_n, d_n)$ of a DHPIM sequence is composed of a header $h_n$, which initiates the symbol, and information slots $d_n$. Depending on the most significant bit (MSB) of the input code word, two different headers are considered $H_0$ and $H_1$. If the MSB of the binary input word is equal to 0, then $H_0$ is used with $d$ representing the decimal value of the input binary word. However, if MSB = 1 then $H_1$ is used with $d$ being equal to the decimal value of the 1's complement of the input binary word. For $H_0$ and $H_1$, the pulse duration is $\alpha T_s/2$ and $\alpha T_s$, respectively, where $\alpha > 0$ is an integer and $T_s$ is the slot duration which is composed of a pulse and guard band. The average symbol length can be reduced by a proper selection of $\alpha$, therefore, DHPIM offers an improved transmission rate and bandwidth requirements.

Adaptive modulation has also been used in order to improve the spectral efficiency and robustness of the channel in space optical links [279], [294], [295]. The efficiency of the adaptive modulation depends upon the accuracy of channel estimates derived from the received signal strength through a low rate feedback path. It makes use of large channel coherence time in millisecond range (due to slowly varying channel) to estimate the state of channel and send it back to the transmitter for varying some of the transmission parameters such as power, coding rate, modulation levels, etc, according to channel conditions. For space-based optical links, adaptive coding is performed where encoder adjusts its code rate to suit slowly varying channel conditions with the help of Rateless codes: (i) punctured codes (puncturing the parity bits to increase the effective code rate) [296] or (ii) fountain codes (in particular Raptor codes where length of the code word is varied to change the rate) [297]. The performance of three different families of rateless codes (Luby Transform, Raptor and RaptorQ) is carried out in [298] for satellite-to-ground FSO links. Short length rateless raptor codes are studied in [299] for ground-to-UAV FSO link with severe jitter. Other adaptive modulation schemes like variable rate, variable power adaptation, or channel inversion can be adopted in case of weak turbulence regime [294]. For strong turbulence regime, truncated channel inversion scheme [300] is adopted where a variable data rate adaptation is performed according to the FSO channel condition. In [301], adaptive data rate is demonstrated for LEO FSO downlink and a throughput gain upto a factor of three is obtained in comparison to constant data rate throughout. When the FSO conditions are favorable, signal constellation size is increased, or when channel conditions are not favorable, signal constellation size is decreased or when intensity channel coefficients are below threshold, the signal is not transmitted at all.

Optical sub-carrier intensity modulation (SIM) is another modulation format where the base-band signal modulates the electrical RF sub-carrier (can be either analog or digital) which is subsequently intensity-modulated by the optical carrier. Since the sub-carrier signal is sinusoidal signal, therefore a DC bias is added to omit negative amplitude of the transmitted optical signal. SIM does not require adaptive threshold unlike OOK scheme and is more bandwidth-efficient than the PPM scheme. Optical

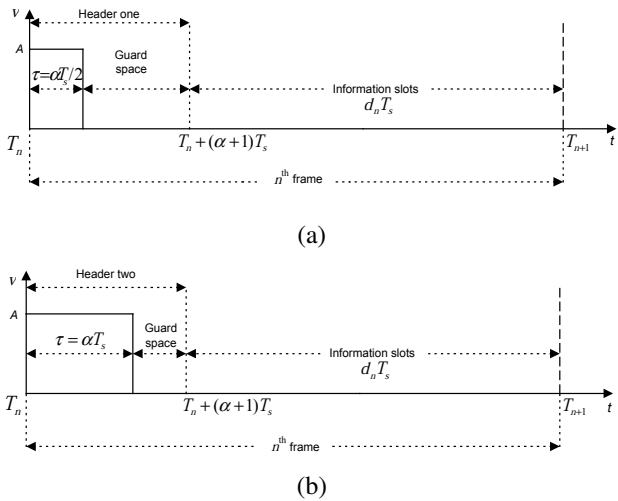

Figure 10. Symbol structure of DHPIM with (a) $H_0$ and (b) $H_1$ headers [293]

A comparison of bandwidth requirement, PAPR and capacity for variants of PPM modulation schemes is shown in Table VII.

SIM inherits the benefits from more mature RF systems, therefore, it makes the implementation process simpler [302]. SIM in conjunction with diversity technique improves the BER performance of the FSO system in the presence of an atmospheric turbulence [303]. When this modulation scheme is used with different RF sub-carriers which are frequency-multiplexed, then this scheme is know as multiple sub-carrier intensity modulation (MSIM). In this case, each sub-carrier is a narrow band signal and experiences less distortion due to inter symbol interference at high data rates. However, the major disadvantage of SIM and MSIM is less power efficiency than OOK or PPM.

Homodyne BPSK is preferred candidate for inter-satellite optical communication links due to its increased sensitivity for both communication and tracking. It also provides full immunity against solar background noise and interferences [304]. The effect of atmosphere on homodyne BPSK communication for satellite-to-ground links is investigated in [304] where it proves to be the robust modulation scheme even in the presence of strong background light. A full duplex 5.6 Gbps optical inter-satellite link using homodyne BPSK in studied in [305]. Another coherent modulation scheme i.e., differential phase shift keying (DPSK) has received significant interest due to its power efficiency and 3 dB improvement over the OOK modulation [306]–[309]. Since it has reduced power requirement than OOK, DPSK does not have non-linear effects which in turn improves the spectral efficiency of DPSK over OOK modulation. It is reported that sensitivity of DPSK receiver can approach the quantum limit theory. However, the cost for implementing DPSK-based FSO systems is high due to its increased complexity in system design both at transmitter and receiver level.

(VI) **Coding:** Error control coding also improves the performance of the FSO link by making use of different FEC schemes including Reed-Solomon (RS) codes, Turbo codes, convolutional codes, trellis-coded modulation (TCM) and LDPC. The study of error performance using error correction codes in fading channel has been under research for many years [310]–[313]. These codes add redundant information to the transmitted message so that any kind of error due to channel fading can be detected and corrected at the receiver. In [314], RS code is proposed for noiseless Poisson PPM-based channel. The performance of RS (262143, 157285, 65) coded 64 PPM provides a coding gain of 5.23 dB over the uncoded system [315]. RS codes provide good coding gain when implemented with PPM for a long distance FSO communication in [316]. RS coding is used along with 4096-ary PPM for laser communication experiment from Earth to Lunar reconnaissance orbiter in the presence of atmosphere turbulence and pointing jitter [317]. In case of strong atmospheric turbulence, Turbo, TCM or LDPC codes are preferred. Turbo codes can be arranged in any of the three different configurations- parallel concatenated convolutional codes, serial concatenated convolutional codes and hybrid concatenated convolutional codes [310], [318], [319]. Parallel concatenated convolutional codes are most popularly used in which two or more constituent systematic recursive convolutional encoders are linked through an interleaver. For very high data rate transmission, LDPC codes are preferred over Turbo codes due to their reduced decoding complexity and computational time. Variable rate LDPC codes can further increase the channel capacity and provide good coding gain [320]. It was observed that the LDPC coded MIMO FSO system using $M$-PPM provides better performance over the uncoded system in case of a strong atmospheric turbulence and large background noise set to -170 dBJ. A coding gain of 10 - 20 dB was observed over the uncoded system at BER = $10^{-12}$ [321], [322]. A comparison of serial concatenated PPM is made with LDPC coded PPM which shows low latency and improved performance for LDPC coded PPM in space-based optical links [323]. Also, bit interleaved coded modulation (BICM) scheme proposed by I.B Djordjevic [321] provides an excellent coding gain when used with the LDPC coded FSO system. Orthogonal frequency division multiplexing (OFDM) combined with suitable error control coding is also considered a very good modulation format for improving BER performance of FSO IM/DD systems. A comparison of different families of Rateless codes i.e., Luby Transform, Raptor and RaptorQ is studied in [298] for satellite-to-ground FSO link and RaptorQ is considered to be the best out of three in order to mitigate the errors in FSO link.

At the receiver, various efficient decoding algorithms have been proposed to decode the generated codes. Theoretically, ML decoding at the receiver can provide better data recovery but its usage is limited due to the implementation complexity. Symbol-by-symbol maximum-a-posterior (MAP) decoding algorithm is computationally complex and is not a preferred choice for implemention on a very large scale integration (VLSI) chip. However, logarithmic version of the MAP (log-MAP) algorithm [324] and the soft output Viterbi algorithm (SOVA) [325] are the practical decoding algorithms for implementation using Turbo codes. Out of these two, log-MAP algorithm gives the best performance but is computationally very complex. Simplified-log-MAP algorithm performs very close to the log-MAP and is less complex aswell [326].

(VII) **Jitter Isolation and Rejection**: For achieving micro-radian pointing accuracy in the presence of platform jitter, a dedicated pointing control subsystem is required that ensures isolation and rejection of platform jitter or satellite vibrations. The mispointing loss of the optical beam without proper rejection of platform jitter can be of the order of more than 10 $\mu$rad depending on the vibration power spectra density. This jitter rejection is accomplished by making use of vibration isolators or compensation control loops [327]. Two basic kinds of isolators are used: passive and active. Passive isolators are mechanical low pass filters that provide a cost-effective and reliable solution for broadband high frequency vibrations. For low frequency vibrations, active isolators are preferred that make use of vibration-control systems, force actuators, and displacement sensors. With the development of smart sensors, actuators and powerful processors, active vibration isolators have gained popularity these days. Various active control loop techniques for jitter rejection include classical feedback, modern feedback, disturbance accommodating control, disturbance observers, repetitive control, adaptive control, adaptive inverse control, adaptive feed forward control, and neural network [328]. The design of these control loops depends on the magnitude and frequency content of the vibrations. In [201], a jitter rejection using self-tuning feed-forward compensation, is developed and it shows that the feed-forward compensator provides a significant improvement over the closed loop control. Other techniques used for jitter control of optical beam are (i) using linear-quadratic Gaussian (LQG) design with feed-forward from accelerometers to fast steering mirrors (FSM) as well as feedback to the FSM using a position sensing detector at the target [329] (ii) using self-tuning regulator (STR) and a filtered-$X$ least mean square (FXLMS) controller that adopts adaptive feed-forward vibration compensation [330] (iii) using combination of least means squares/adaptive bias filter and linear-quadratic regulator for controlling sinusoidal and random jitter [331]. Adaptive methods such as the least means squares (LMS), broadband feed-forward active noise control, model reference, or adaptive lattice filters are also used to control jitter. LMS-derived methods require prior knowledge of the system dynamics, which may vary with time. An implementation of various vibration isolation algorithms such as the multiple LMS algorithm, the clear box algorithm, and the adaptive disturbance canceller have been carried out experimentally using hexapod platforms in [332]. It was observed that multiple LMS requires a seperate measured disturbance-correlated reference signal in order to perform vibration compensation. It is an efficient computational algorithm when disturbance-correlated signals are available. The performance of the clear box algorithm implemented either in time domain or frequency domain exceeds that of multiple LMS algorithm without requiring a measured disturbance-correlated signal. The time domain clear box sine/cosine method is preferred when the disturbance frequencies are time-independent or just varying slowly, but in case of rapidly varying disturbance frequencies, the adaptive basis version should be used as it provides a better performance. Adaptive disturbance canceller works well in suppressing static disturbance frequencies however, it requires an additional frequency identification algorithm and disturbance measuring devices. In [333], a fiber based coherent receiver along with fine pointing subsystem is implemented to minimize the effect of satellite platform vibrations. They demonstrated the receiver sensitivity of -38 dBm at the transmission rate of 22.4 Gbps for polarization-multiplexed QPSK signal in inter-satellite link.

Another cause for jitter in the receiver detector plane is the atmospheric turbulence-induced random angle-of-arrival fluctuations. Turbulence-induced jitter effects can be greatly improved having a receiver tip-tilt correction system that includes two-axis FSM located at the image plane in order to compensate for the atmospheric turbulence-induced angle-of-arrival fluctuations. In [334], a piezoelectric FSM is implemented on a tracking test bed with bandwidth up to 1 kHz. The tracking feedback loop (TFL) was characterized by finding the frequency response of the gimbal tracking interface and summing the logarithmic magnitude and phase response plots with those of the FSM. A significant reduction in jitter was observed using TFL. Besides using FSM, a liquid crystal beam steering device along with non-linear adaptive controller can also reduce jitter fluctuations [335]. It increases the durability of the tilt corrector as it does not involve mechanical moving part and reduces the power requirement of the controller due to its low voltage operation.

(VIII) **Background Noise Rejection:** The major source of background noise is due to day-time solar radiations. The amount of day-time background noise is strongly dependent on operating wavelength, with lesser background noise at higher wavelength. It can be mitigated with the help of spatial filters along with a suitable modulation technique that has a high peak-to-average power. Important design considerations in selecting the filter are the angle-of-arrival of the signal, the Doppler-shifted laser line width, and the number of temporal modes [280]. The most suitable modulation scheme is $M$-PPM to combat the effect of solar background noise radiations (as the noise is directly proportional to the slot width) [155]. High order PPM scheme is reported as a potential modulation scheme for inter-satellite links as it is more power efficient and drastically reduce the solar background noise [336]. Designing receiver with narrow FOV and choosing filters with spectral width less than 1 nm is another approach to reduce the background noise [337]. It has been shown that adaptive optics and deformable mirrors consisting of an array of actuators can result in a significant improvement due to the background noise by reducing the receiver FOV. The analysis presented in [338] showed that the inter-planetary FSO link between Earth and Mars has achieved 8.5 dB improvement in extreme background and turbulent conditions using adaptive optics and array of actuators with PPM modulation scheme. During moderate background conditions, the improvement was decreased to 5.6 dB. Another analysis presented in [339] gives the performance improvement of 6 dB by using array of 900 actuators and adaptive optic technique with the 16 PPM modulation scheme. A coherent detection scheme employing LO provides improved sensitivity and low susceptibility to background radiations. However, coherent detection scheme, either homodyne (i.e., optical signal is directly transferred into the baseband) or heterodyne (i.e., there is a frequency difference between LO and signal) requires perfect phase alignment of the received and LO signal which leads to increase in the complexity of the receiver. The effect of background noise and phase errors on coherent optical receiver for FSO channel was studied in [172]. Significant suppression of background radiations have been observed in case of optical homodyne receivers proving their ability to achieve quantum limit performance in a high background environment. In [340], strong background solar radiations are reduced using a heterodyne system with a very narrow intermediate frequency (IF) and therefore, permits the usage of the FSO system even when the Sun is directly in FOV without any degradation.

(IX) **Hybrid RF/FSO:** The performance of the FSO communication is drastically affected by weather conditions and atmospheric turbulence. This can lead to link failures or poor BER performance of the FSO system. Therefore, in order to improve the reliability and improve the availability of the link, it is wise to pair the FSO system with a more reliable RF system. Such systems are called hybrid RF/FSO and are capable of providing high link availability even in adverse weather conditions [341], [342]. The major cause of signal degradation in RF transmission is the rain (as the carrier wavelength is comparable to the size of the rain drop) and in the FSO communication is due to fog. So, the overall system availability can be improved by using a low data rate RF link as a back up when the FSO link is down. In [343], the availability of an airborne hybrid RF/FSO link is evaluated. It is observed that the FSO link provides poor availability during low clouds conditions due to the attenuation by cloud particles and temporal dispersion. A significant improvement is observed when an hybrid RF/FSO link is used as RF signals are immune to cloud interference. However, the conventional approach of hybrid RF/FSO causes inefficient use of channel bandwidth. Also, a continuous hard switching between the RF and FSO systems could bring down the entire system. Therefore, a new approach as suggested in [344] gives the symbol rate an adaptive joint coding scheme wherein both the RF and FSO subsystems are active simultaneously which saves channel bandwidth. Also, rateless coding approach provides a performance advantage over fixed or adaptive rate coding scheme regardless of channel conditions [296], [345], [346]. These codes do not require channel knowledge at the transmitter and automatically adapts the rate between RF and FSO links depending on weather conditions with single bit feedback per message. The hybrid RF/FSO link provides great application for airborne vehicles [343], [347]. This technology is also used to improve the capacity and interference problem in conventional RF communication [348]–[350].

### B. TCP Upper Layer Methods

There has been a lot of research on the performance mitigation of the atmospheric turbulence in physical layer. For the last few years, researchers have gained attention to work on modeling and performance evaluation of other layers including link layer, network layer or transport layer in order to improve the performance of the FSO communication [351]–[353]. In addition to the physical layer methods, various techniques like re-transmission, re-routing, cross connection between different layers, delay tolerant networking, etc., are used to improve the performance of FSO in all weather conditions [354], [355].

(I) **Re-transmission:** A re-transmission protocol such as automatic repeat request (ARQ) is widely used in data communication for reliable data transfer [356]. Here, the transmission is carried out in the form of packets of certain frame lengths. If due to some reason, the receiver does not acknowledge the transmitted packet within the speculated time frame, then the packet is re-transmitted. This process repeats until positive acknowledgment is received by the transmitter from the receiver or the preset counter value is exceeded. So, this kind of stop, wait and go-back-$N$ ARQ scheme results in a huge delay, large energy consumption and bandwidth penalties due to re-transmission process [357]. Therefore, another variant of ARQ is selective repeat ARQ (SR-ARQ) in which data packets are continuously transmitted from the transmitter to the receiver without the need to wait for individual acknowledgment from the receiver. The receiver will continue to accept and acknowledge the received frame. If any frame is not acknowledged after certain period, it is assumed to be lost and re-transmitted. ARQ protocol can be implemented either at data link layer or at transport layer [358]. In either case, the receiver terminal must have sufficient data storage capability to buffer the received data at least for the time period specified by the window size. In [359], the performance of inter-HAP optical link in terms of optical efficiency is studied using feedback information in ARQ data link layer protocol.

Another variant of ARQ that has been studied by various researchers is hybrid-ARQ (H-ARQ) which uses a combination of FEC coding and ARQ error control [360]–[362]. The performance of FSO downlink from satellite-to-ground with ARQ and FEC schemes is studied in [363], [364]. The outage probability of different H-ARQ scheme in the strong turbulence regimes is investigated and it is found that good performance gain is achievable using this scheme [365]. However, H-ARQ scheme has large bandwidth penalties and delay latencies. Recently, a combination of cooperative diversity with ARQ (C-ARQ) has gained interest that has shown remarkable results for combating turbulence-induced fading in the FSO channel. Another version of C-ARQ with lesser transmission delays and improved energy consumption is the modified cooperative diversity with ARQ (MC-ARQ). This modified scheme allows relay nodes to store a copy of frames for a more efficient response during transmission failure due to atmospheric turbulence [366], [367]. Another protocol i.e., Rateless Round Robin protocol is used to provide reliable FSO communication in case of significant outrages. [368], [369]. It is observed that Rateless Round Robin is an effective error control design for practical FSO applications even during very strong turbulence when the channel availability is less than 45% [370].

(II) **Reconfiguration and re-routing:** Path reconfiguration and data re-routing is carried out in order to increase the availability and reliability of the FSO link during loss of LOS or adverse atmospheric conditions or device failure. Through dynamic reconfiguration of the nodes in the FSO network using physical and logical control mechanism, link availability is improved drastically. In physical layer, the reconfiguration is achieved using ATP and in logical layer, it makes use of autonomous reconfiguration algorithms and heuristics. Here, the data packets are re-routed through other existing links that could be either an optical link or low data rate RF link. Autonomous reconfiguration is used in optical satellite networks where the link is dynamically routed to alternate path if there is any link failure in the current path [371]. An automated ATP technique using dynamic path reconfiguration is present in [348], [372]. The dynamic path reconfiguration in hybrid RF/FSO network is demonstrated experimentally in [373]. Various topology control mechanisms and re-routing algorithms have been investigated for the FSO network [374]–[378]. Topology control and beam reconfiguration are achieved through: (a) the topology discovery and monitoring process, (b) the decision making process by which a topology change has to be made, (c) the dynamic and autonomous re-direction of beams (based on algorithms) to new receiver nodes in the network, and (d) the dynamic control of these beams for link re-direction [374]. Therefore, reconfiguration and re-routing improve the reliability of the FSO link but at the cost of a huge processing delays. A good design engineer has to ensure the restoration of the link through reconfigurability without significant impact in delay and at reduce cost. For this, the routing protocol should be designed in such a way so that during re-routing process, the path which has minimum delay or least number of hops should be given priority. Sometimes, all the routing routes are computed prior to their actual need and are stored in routing tables. Such type of routing is classified as 'proactive routing' protocol. However, this routing protocol is not suitable for large networks as it imposes high overhead to the network which makes it bandwidth inefficient. Another routing protocol that generates very less overhead as compare to proactive routing is called 'reactive routing' and computes new routes only 'on demand'. A new route is established only during the failure of the existing route. However, this leads to prolonged latency in data delivery. Combination of both proactive and reactive routing protocols is called 'hybrid routing' protocol. It divides the network into clusters and apply proactive route updates within each cluster and reactive routing across different clusters.

(III) **Quality of Service Control:** The QoS in the FSO communication is measured in terms of data rate, latency, delay jitter, data loss, energy consumption, reliability and throughput efficiency. The data transfer from one node to another in the FSO communication system should meet the given requirements of special QoS class otherwise the offered services can not be used by end-users in a satisfying way. For this, the main challenge in the FSO network is to optimize the performance of the communication system measured in (a) end-to-end connection delay, (b) delay variation, (c) packet rejection rate, and (d) overhead. A quality-of-service (QoS) multicast routing protocol is studied for HAP-satellite link in [379]. A 10 Gbps QoS-based buffer is proposed in [380] in order to mitigate packet loss during strong turbulent conditions in the hybrid RF/FSO network. The buffer incorporates a custom IP packet inspection and scheduling processor which works in accordance with the link availability and QoS parameters. It was shown that the FSO link using this buffer can provide link availability under strong atmospheric conditions at fade margins as low as 8 dB. It is suggested in some literatures that modification in different layers improves the QoS of FSO system. The QoS requirements set up by International Telecommunication Union (ITU) and 3rd Generation Partnership Project (3GPP) for various FSO scenarios (satellite-to-ground, aircraft-to-aircraft, maritime mobile, etc.) is studied in [381]. A routing algorithm that improves the QoS of the network and the medium access control (MAC) layer is proposed in [382]. The MAC layer QoS can be classified into (a) channel access policies, (b) scheduling and buffer management, and (c) error control. This routing protocol provides an energy efficient real time FSO communication.

(IV) **Others:** Re-playing is another technique to promote end-to-end connectivity of the FSO link. If re-routing or re-transmission is not possible, then the FSO network will replay up to 5 sec of data from the edge node [383]. Delay (or disruption) tolerant networking (DTN) technique is applied for the networks with intermittent connectivity and therefore, is a good candidate for the FSO communication having extreme atmospheric conditions [384], [385].

## V. ORBITAL ANGULAR MOMENTUM FOR THE FSO SYSTEM

Angular momentum is one of the most fundamental physical quantity in both classical and quantum mechanics. It is classified as spin angular momentum (SAM) and orbital angular momentum (OAM). SAM is associated with the spin of the photon and thus, it is related with polarization. On the other hand, OAM is associated with the helicity photon wavefront and therefore, it is related to spatial distribution. Recently, researches have directed their attention for exploiting the capabilities of OAM beam in the FSO communication system [311], [386]. It has been seen that the FSO communication has not fully utilize the Terabit capacity of the optical carrier. In order to fully explore the potential of the FSO system, lot of research is going on in the area of OAM-multiplexed free space communication link. Information carrying OAM beams are capable of providing an increased capacity and spectral efficiency. Therefore, this approach is considered as a preferred choice for achieving high data returns from deep space or near Earth optical communication in near future [311]. At present, power efficient PPM scheme is most widely used for the long distance or deep space communication. However, the need to achieve high data rates for future space links makes the PPM a weak candidate due to its low spectral efficiency. Also, the usage of large number of slots in PPM for meeting high capacity demands increases the complexity and cost of the system. In this case, OAM modulation is considered to satisfy a high bandwidth requirements for future space missions while keeping the system cost and power consumption reasonably low. The main challenge in this approach is to preserve orthogonality between OAM states in the presence of the atmospheric turbulence. However, OAM beams when used with suitable coding, MIMO or equalization techniques are capable of operating even under strong atmospheric turbulence regime [311], [387], [388].

It was reported in [389] that a beam having helical shaped-phase front described by azimuthal phase term $\exp(il\theta)$ carries OAM of $l\bar{h}$ per photon where $l$ is topological charge with any integer value, $\theta$ is the azimuth angle and $\bar{h}$ is the Plank's constant $h$ divided by $2\pi$. Therefore, unlike SAM, which can take only two possible states of $\pm\bar{h}$, the OAM can have an infinite number of states corresponding to different values of $l$. In principle, infinite number of bits are carried by OAM of single photon. This makes OAM a potential candidate for high capacity communication systems. Also, orthogonality among beams with different OAM states allow additional degree of freedom by multiplexing of information carrying OAM beams. The possibility to generate and analyze states with different OAM by using interferometric or holographic methods [390]–[392] permits the development of energy-efficient FSO communication systems. Further, the OAM-based FSO system has provided a good performance in atmospheric turbulence when used with suitable encoding or modulation format or adaptive optics system.

An OAM beam is formed by attaching azimuthal phase term to the Gaussian beam as $U(r,\theta) = A(r) \cdot \exp(il\theta)$. Here, $A(r)$ is the amplitude at the waist of the Gaussian beam and $r$ the radial distance from the center axis of the Gaussian beam. When data is encoded to the OAM beam, it is expressed as $U(r,\theta,t) = S(t) \cdot A(r) \cdot \exp(il\theta)$, where $S(t)$ is the data to be transmitted. With multiplexing of $N$ information carrying OAM beams, the resultant field is expressed as $U_{\text{Mux}}(r,\theta,t) = \sum_{m=1}^{N} S_m(t) \cdot A_m(r) \cdot \exp(il_m\theta)$. It is to be noted that the OAM of an individual beam is not modified when propagated through free space or spherical lenses. For de-multiplexing OAM beams, an inverse of azimuthal phase term $\exp[i(-l_n)\theta]$ is used and received de-multiplexed beam is then given as

$$U_{\text{Rx}}(r,\theta,t) = \exp[i(-l_n)\theta] \cdot \sum_{m=1}^{N} S_m(t) \cdot A_m(r) \cdot \exp(il_m\theta). \quad (29)$$

The capacity and spectral efficiency of FSO links using OAM is increased by employing several techniques like: (i) combining multiple beams with different OAM values, (ii) using positive or negative OAM values, (iii) using wavelength division multiplexing (WDM), (iv) polarization multiplexing, or (v) using two groups of concentric rings. Recent reports in [393] have demonstrated a 2.56 Tbps data rate transmission with spectral efficiency of 95.7 bps/Hz using four light beams with 32 OAM modes employing 16 QAM. In [394], a 100.8 Tbps data transmission has been reported using 42 wavelengths with 24 modes. This shows that OAM has tremendous potential for increasing the capacity of the FSO system. However, these high capacity transmissions were limited to short transmission distances only where the effect of turbulence was not considered. Therefore, OAM-based FSO system is capable of delivering huge data returns in inter-satellite links or deep space mission where the atmospheric turbulence does not pose any problem. A multi-gigabit transmission (projected for 2020) may use coded OAM modulated beams to satisfy high bandwidth demands for future deep-space and near-Earth optical communications [334].

It has been reported in [395] that OAM beams are highly sensitive even in case of a weak atmospheric turbulence due to the redistribution of energy among various OAM states leading to time-varying crosstalk. An OAM beam has a doughnut shape with less power and large phase fluctuations in the center. Since the orthogonality of the beam is dependent on the helical phase front, therefore practical implementation of OAM-based FSO system in the presence of atmospheric turbulence is a challenging task. Single information carrying OAM state results in a random and bursty error in the presence of atmospheric turbulence. In case of multiplexed data channels having different OAM values, the crosstalk among adjacent channels degrade the performance of the system [396]. In [397], an experimental investigation was carried out for the OAM-based multiplexed FSO communication link through an emulated atmospheric turbulence. The results indicated that turbulence-induced signal fading and crosstalk could significantly deteriorate the link performance. However, very recently, researchers have reported that the combination of suitable error correcting codes and wavefront correction techniques has shown good results in mitigating the effect of turbulence for the OAM-based FSO system. Channel codes are used to correct the random errors caused by the atmospheric turbulence for single OAM state and wavefront correctors take care of cross talk among adjacent OAM states. Various channel coding techniques like LDPC codes in [398], RS codes in [399] have proved quite beneficial improvements in the performance of OAM-based FSO system. LDPC codes when used with OAM-based modulation schemes are capable of operating even under strong atmospheric turbulence regime [311]. In [400], RS codes in combination with Shark-Hartmann wavefront correction method have been adopted to combat the detrimental effect of atmospheric turbulence. It was shown that the effect of atmospheric turbulence was reduced and satisfactory BER performance was achieved even during strong atmospheric turbulence. Other techniques like the holographic ghost imaging system [401], adaptive optics [402], [403] help in controlling OAM crosstalk and thus are capable of providing high data rates even in adverse atmospheric conditions. In [404], 80 Gbps FSO link is experimentally demonstrated using OAM and MIMO-based spatial multiplexing. For the practical implementation of OAM-based FSO multiplexed system, various design considerations such as beam size, receiver aperture size, mode space of OAM beams for given lateral displacements are analyzed in [405]. Fig. 11 shows the block diagram representation for the OAM-based FSO multiplexed system using channel coding technique. The transmitter consists of encoder, modulator, laser source, computer controlled hologram along with spatial light modulator (SLM) and multiplexer. The data from different independent sources are encoded, modulated and transformed into OAM beams by adding a spiral phase mask with different charges ($l$). These OAM beams are then multiplexed together and sent onto the FSO channel. The multiplexing of OAM beams is considered as a form of spatial multiplexing which is capable of enhancing the capacity and the spectral efficiency of the FSO link. At the receiver side, OAM beams are de-multiplexed and an inverse spiral phase mask with charge ($-l$) is used to remove the azimuthal phase term $\exp(il\theta)$ of OAM beam to recover the plane phase front of the beam. This optical beam is then passed through detector, demodulator and decoder to recover information data.

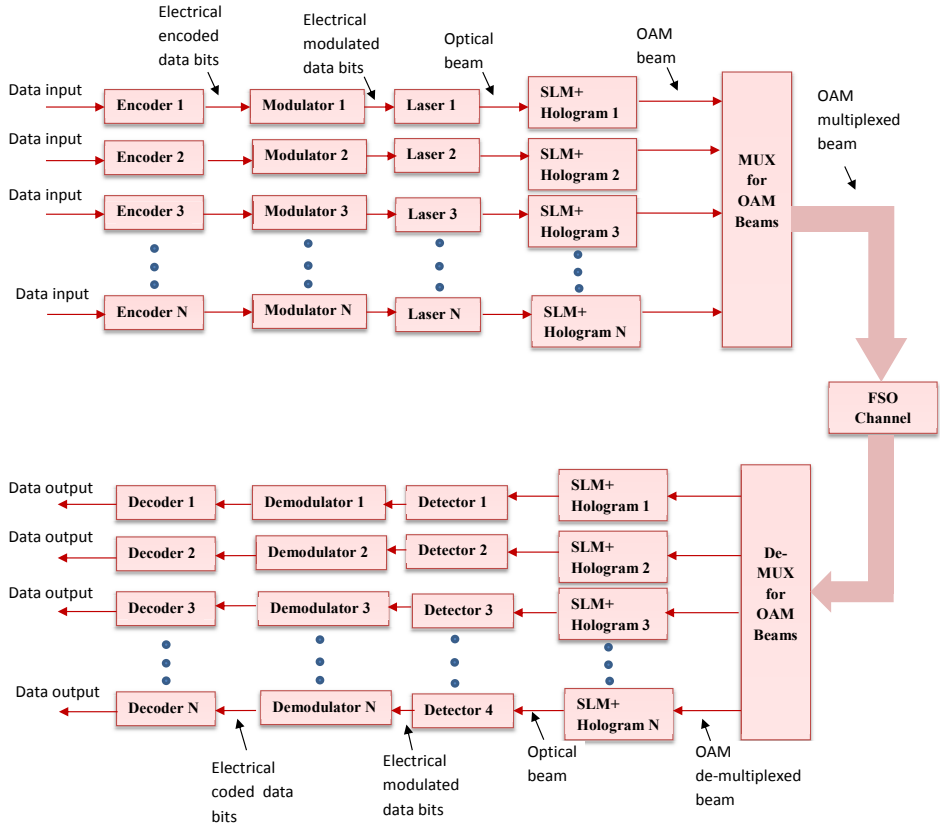

Figure 11. FSO communication system using multiplexed OAM beams and channel coding

## VI. FSO BACKHAUL COMMUNICATION

The demand for high date rate is increasing in today's network and researchers these days are finding various techniques to provide gigabit capacity low cost backhaul solutions [406]–[409]. Space-based links are generally satellite or UAV links usually used to provide high speed connections where terrestrial links are not available. It has been reported in [410], that the mobile backhaul in United States will increase 9.7 times by 2016 due to rapidly growing data demands. This significant capacity growth has demanded the need for reconfigurable broadband communication networks with high data rates and reasonable cost. Further, the increase in mobile traffic has led to large number of cell sites which demands more bandwidth and large number of backhaul connections per cell site. Conventional technologies used for backhaul network include either copper lines, optical fiber, or RF backhaul. Copper lines are becoming an infeasible option for meeting future backhaul demands as they have a low data rate and their prices increases linearly with capacity. Fiber optics seems to be a good choice for high capacity links over long distances, however, they require considerable deployment time and huge initial investment. In that case, radio waves can be used but their capacity taps out around 400 Mbps due to the limited RF channel bandwidth. Although RF has played a vital role in backhaul for 2G and 3G networks for many years, however, huge data rate requirements for next generation mobile networks (5G), drive the need for high capacity backhaul links [411]. In order to cope up with increasing data rate, multiple signals using multiple RF can be transmitted but that will lead to additional electronics and an additional spectrum licensing cost. As frequencies goes up to milliwave range (60/80 GHz), high capacity links for short distance applications can potentially be employed. However, they require a very large spectrum allocation (10 GHz) to approach even 1Gbps. Further, they demand a very tight beam alignment and have reliability challenges due to rain fades.

The FSO communication has recently gained researchers attention to provide gigabit capacity backhaul links due to its low cost and rapid deployment speed in comparison with conventional backhaul technologies like RF or optical fibers [69], [412]. This survey focuses on optical backhaul links between HAPs, satellites and ground-based receivers. The optical backhaul communication between HAPs, satellite and ground are considered to serve as future broadband backhaul communication channels [413], [414]. HAPs are quasi-stationary vehicles like helium filled airships or air-crafts placed in the stratosphere, at an altitude 17-25 km where the impact of the atmosphere on optical beam is less severe than directly above the ground. Since the HAPs are located in a cloud-free atmospheric altitude, they are capable of providing reliable links between different HAPs or between HAPs and air-crafts or between HAPs and satellites. HAPs has unique characteristics compared with terrestrial and satellite systems such as large coverage area (3-7 kms), quick deployment, flexible capacity increase through spot beam re-sizing, low maintenance cost and broadband capability. Since the HAPs are placed far from the the atmospheric region, they provide a better channel conditions than satellites. Moreover, they provide better LOS condition in almost all coverage areas, thereby causing less shadowing than terrestrial systems. They can act as a relay station to forward the high capacity optical data through the atmosphere to the ground. The backhaul optical link for the HAPs is capable of connecting to the core network through terrestrial gateway stations. In case whenever HAP network is much larger than the cloud coverage correlation length, a ground station diversity can be used to improve the reliability of the system [415]. A multi-beam antenna is required for covering large subscriber ground stations by single HAPS with high frequency reuse efficiency. Furthermore, employing on-board regenerative HAP payload can split the satellite-to-ground link into two parts: (i) satellite-to-HAP link where the attenuation is equal to free space loss and (ii) HAP-to-ground link which is affected by the atmospheric attenuation. This would not only provide high capacity backhaul optical links but also relax the satellite front end requirements and reduce its on-board processing time. HAPs can work either as stand alone system or can be integrated with other satellite or terrestrial system [416] as shown in Fig. 12. The integrated ground-HAP-satellite system provides broadcasting and broadband services over a wider coverage area. It can be used to provide services to sub-urban areas with very little cost of deployment. Multiple optical payloads embarked in each HAP can serve as a large solid-state memories (almost in Terabyte size) which can store the collected data from the satellite and forward it to the ground station at any time which is not constrained by the satellite visibility time. HAPs in a stand-alone system can be deployed economically for providing high capacity and wider coverage to rural or remote areas.

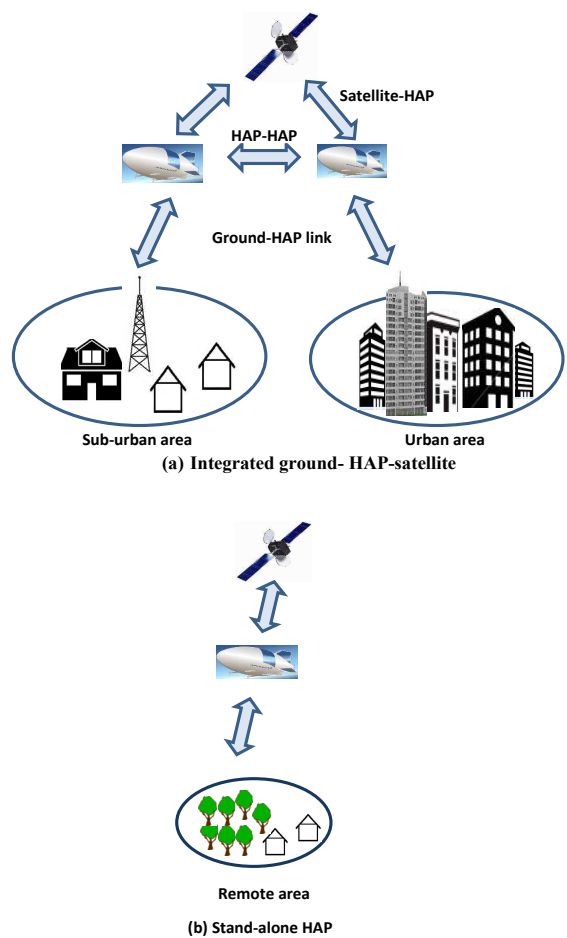

Figure 12. HAP system architecture

A high bit rate optical backhaul downlink was demonstrated using HAP as a part of European "CAPANINA" (communications from aerial platform networks delivering broadband communications for all) project [417]–[419]. This was the first FSO link trail from the stratosphere. In this stratospheric optical payload experiment (STROPEX), a free space experimental laser terminal (FELT) was mounted at approximately 22 kms and a high capacity downlink from FELT was observed to transportable optical ground station (TOGS) as shown in Fig. 13. FELT was able to transmit at three different rates: 270 Mbps, 622 Mbps and 1.25 Gbps. Using IM/DD and standard APD detector, the sensitivity of the receiver front end was measured at 168 photons per bit.

STROPEX also focused on the ATP system to ensure good tracking capability in stratospheric conditions.

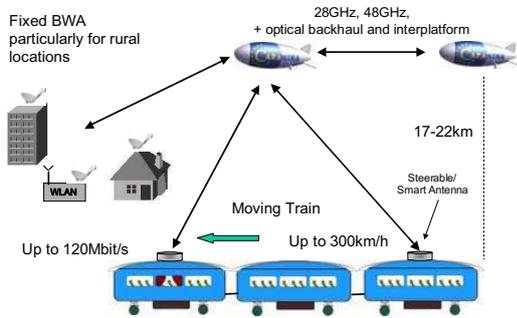

Figure 13. Demonstration of optical backhaul link for CAPANINA [420]

A theoretical analysis for horizontal link between two stratospheric HAPs at 20 kms in [421], [422] showed that a high capacity link is feasible upto almost 600 kms using intensity modulation at 1550 nm. The system allows for a BER of $10^{-6}$ at a data rate of 384 Mbps and a transmit power of 800 mW. The effect of the atmospheric turbulence on the performance of the HAP-satellite link has been studied in [423] and it was shown that an optical backhaul link between HAP and GEO satellite can be established upto 10.7 Gbps using return-to-zero (RZ) intensity modulation in combination with FEC. The inter-networking between HAPs and LEO satellites deals with physical layer problems such as managing Doppler shift, ATP, optical transmitter design, etc. A HAP-GEO [424] and HAP-LEO [425] optical links using DPSK modulation scheme offers 3 dB performance improvement over OOK. The system design requirements from HAP-LEO link is quite different from HAP-GEO link. In [426], system design requirements for optical communication between HAPs and satellite network is presented taking into account various critical aspects such as low elevation links, the effect of Sun, propagation effects, etc. An optimization problem between HAPs and non-geostationary (NGEO) mobile satellite systems is proposed in [427] for maximizing the utilization of HAPs and the average elevation angle between HAPs and satellites. They have also proposed a polynomial-time solution approach for avoiding frequent switching of optical links with an appropriate selection of system parameters. In [428], a concept of optical transport network based on HAPs employing dense wavelength division multiplexing (DWDM) has been investigated. Taking into account the physical constraints imposed by the FSO, they have estimated the number of a wavelengths required for full inter-connectivity without wavelength conversions for all optical routing in the network. Analysis in [429] showed a poor correlation among the signals in case of wavelength diversified ground-HAP FSO link in the presence of radiation fog.

The effect of laser non-linearity and atmospheric turbulence on the performance of optical HAP-ground link has been analyzed using hybrid millimeter wave/optical backhaul link [430]. Here, the fronthaul millimeter wave carrier signals use optical frequency division multiplexing (OFDM) which directly modulates the backhaul optical carrier to ensure reliability in all weather conditions and improves the system throughput. OFDM transmissions are characterized by sporadic occurrences of very high peak power and when this signal is applied to the laser, it results in distortions. Use of amplifiers with very large dynamic range or clipping the signal can help in minimizing inter-modulation distortion. The optical backhaul and inter-HAP links when combined with RF carrier provide promising solutions even in the adverse weather condition. Numerous models have been proposed to formulate the problem of backhaul network design using the hybrid RF/FSO technology [416], [431].

The geographical location of HAP in the stratosphere is very critical as the climatic conditions and the stratospheric winds determine the link availability of the system. The positioning accuracy of HAP determined by the wind speed in the stratospheric zone is presented in [432]. This paper also discusses the capabilities of HAP and GEO relay stations to increase the downlink capacities of LEO satellites. The stratospheric wind speeds are more strong around $\pm 60°$ latitude, which complicates station keeping. Therefore, geographic latitudes around $\pm 30°$ are considered most favorable. Table VIII gives some examples of HAPs with their altitude and data rates used in optical communication missions.

| Optical Mission | Data Rate | Altitude | Communication |
|---|---|---|---|
| CAPANINA [420] | 1.25 Gbps | 25 km | HAP-ground |
| Cost 297 (WG2) [433] | 384 Mbps | 20 km | HAP-HAP |
| Helinet [434] | 120 Mbps | 17 km | HAP-HAP |
| ATENNA [435] | 100 Mbps | 18 km | Satellite-HAP and HAP-HAP |

Table VIII
EXAMPLE OF HAPs USED IN OPTICAL COMMUNICATION MISSIONS

## VII. FUTURE SCOPE

The FSO communication has experienced a rapid growth in the last few years despite of various crisis in the global market. This technology has demonstrated less capital expenditure with huge returns in very little time due to (i) easy availability of components, (ii) quick deployment (as it does not seek permission from municipal corporation for digging up of street), and (iii) no licensing fee required. . The FSO technology provides a good solution for cellular carriers using 4G technology to cater their large bandwidth and multimedia requirements by providing a backhaul connection between cell towers. Using ultra short pulse (USP) laser, the FSO communication provides up to 10 Gbps backhaul connection without deploying fiber cables [436]. Also, with the help of advanced modulation schemes, the wireless backhaul capacity can further be increased even beyond 100 Gbps which will suffix the requirement for future 5G cellular networks [437]. The architecture of fifth generation internet system consist of interconnecting satellites with airborne and ground-based receivers via optical links which are supported by terabit wide laser backbone, located at synchronous altitude [438]. In future, instead of hybrid fiber-coax systems, the hybrid fiber-FSO system may cater the high bandwidth and high data rate requirements of end users [439].

The FSO technology allows connectivity to remote places where physical access to 3G or 4G signals is difficult. It involve integration of terrestrial and space networks with the help of HAPs/UAVs by providing last mile connectivity to sensitive areas (e.g., disaster relief, battlefields, etc.) where high bandwidth and accessibility are necessary. An upcoming project by Facebook is an example that will allow internet access to sub-urban or remote areas by providing aerial connectivity to users using the FSO link [440], [441]. It is proposed that for sub-urban areas in limited geographical regions, solar-powered high altitude drones will be used to deliver reliable internet connections via FSO links. For places where deployment of drones is uneconomical or impractical (like in low population density areas), LEO and GEO satellites can be used to provide internet access to the ground using the FSO link.

FSO technology when used over a mobile platform can be deployed in armed forces as it demands secure transmission of information on the battlefield. Intelligence, Surveillance, and Reconnaissance (ISR) platforms can deploy this technology as they require to disseminate large amount of images and videos to the fighting forces, mostly in a real time. Near Earth observing spacecrafts/drones can be used to provide high resolution images of surface contours using synthetic aperture radar (SAR) and light detection and ranging (LIDAR). In case of security monitoring system, airships or helicopters hovering over larger border area would be more economical than constructing walls or employing personnel stretched over thousands of kilometer. In this case, the data collected by the sensor would be transmitted to the command center via on-board satellite communication sub-system using FSO technology. This technology can also be used for space-based detectability and identification of submersibles which requires complete system integration of ground-based, space-based and airborne platforms. In this, a high energy laser pulse is transmitted from a ground-based station towards a space-based mirror that scans over a specific oceanic area looking for returns from underwater vehicles. The returns in blue-green region are then picked up by LEO satellite or airplane or HAP which are then communicated to the command center [442].

Interplanetary optical communication between Moon and Earth or between Mars and Earth are another important missions for NASA. The Lunar Laser Communication Demonstration (LLCD) project by NASA has successfully carried out two way communication using a laser between Moon and Earth. It has demonstrated 20 Mbps uplink (which supported high density video) and record breaking downlink rate of 622 Mbps [443], [444]. This historic success of LLCD's mission has given new impetus to NASA's deep space optical communication. Access links for the Moon and Mars or bidirectional high data rate communication between Mars and Earth are some of the attractive upcoming developments in deep space missions. However, deep space optical communication has additional challenges from LLCD which includes kilowatt class laser beam, a photon-counting detector array on the spacecraft to observe the uplink beam, larger receiver aperture on the ground, larger point ahead angle for downlink beam and stable beam pointing system. The LCRD project by NASA will be the first long period space-based optical communication project which is to be used for both near Earth and deep space missions [37]. It is expected to transmit data at rates 10 to 100 times faster than any RF systems and will be used to deliver high resolution images or videos from various planets over solar system.

Another upcoming technology that supports fifth generation mobile network is the deployment of steered laser transceiver (SALT) on a nanosatellite (or picosatellite) whose weight is assumed to be less than 20 lb (or less than 3 lb). Due to their small size and low launch cost, it is expected to launch about 1000 small satellites into LEO orbit for the purpose of Earth exploration, data imaging, tracking, weather sensing, etc. using FSO technology [445]. The power consumption is assumed to be 0.5 to 5 W depending upon the data rates. SALT or Advanced SALT (ASALT) system would be used for optical communication between aircraft and ground station or inter-satellite communication upto several 1000 kms.

## VIII. CONCLUSIONS

The tremendous growth in the number of multimedia users and internet traffic in the recent years has incurred a substantial strain on the RF system operating at low data rates. Due to this huge explosion in information technology that is driving the information business to higher and higher data rates, there is a need to switch from the RF domain to optical domain. The FSO communication is capable of providing LOS wireless connection between remote sites with very high bandwidths. This technology is considered to be the promising technology in near future which can meet very high speed and huge capacity requirements of current day communication market. However, in order to fully utilize the terabit capacity of the FSO system, it has to overcome various challenges offered by the heterogeneous nature of the atmospheric channel. The FSO system is vulnerable towards various atmospheric phenomena like absorption, scattering, atmospheric turbulence and adverse weather conditions. Various techniques implemented either at physical layer or at other layers (link, network or transport layer) help to combat the detrimental effect of the atmosphere on the quality of the laser beam. Several fading mitigation techniques that were initially proposed for the RF works well for the FSO communication aswell e.g., diversity, adaptive optics, error control codes, modulation, etc. Besides this, the complementary nature of the RF and the FSO communication has motivated the design of an hybrid RF/FSO system which ensure a carrier class availability for almost all weather conditions. Also, modifications in the other TCP layers like transport or network layers with suitable protocols and algorithms help in improving the reliability of the FSO system. The OAM-based FSO system is capable of a high capacity and an increased spectral efficiency in combination with other degree of freedom such as polarization, wavelength multiplexing or multi-amplitude/phase modulation formats.

The OAM-based FSO system has recently gained attention as a potential candidate for the deep space and near-Earth communication system when used with suitable coding or modulation format or adaptive optics techniques to deal with atmospheric turbulence. Further, gigabit capacity backhaul solutions with relatively low cost is achievable using the FSO system. The FSO backhaul downlink up to 1.25 Gbps has been demonstrated experimentally and theoretical studies showed the feasibility up to 10 Gbps for HAP-HAP or HAP-satellite backhaul links.

Hence, it is clear that after so much advancement in the FSO communication, this technology seems to have very high growth prospects in the near future. Many commercial products for FSO links are already available in the market and hopefully very soon this technology will bring worldwide telecommunication revolution.

## List of Acronyms

| | |
|---|---|
| 3GPP | 3rd Generation Partnership Project |
| AFTS | Airborne Flight Test System |
| ALM | Aeronomy Laboratory Model |
| AO | Adaptive Optics |
| ARQ | Automatic Repeat Request |
| ATP | Acquisition, Tracking and Pointing |
| BER | Bit Error Rate |
| BICM | Bit Interleaved Coded Modulation |
| C-ARQ | Cooperative-Diversity with Automatic Repeat Request |
| CALIPSO | Cloud-Aerosol Lidar and Infrared Pathfinder Satellite Observation |
| CAPANINA | Communications from Aerial Platform Networks delivering Broadband Communications for All |
| CCD | Charge Coupled Device |
| CEMERLL | Compensated Earth Moon-Earth Retro-Reflector Laser Link |
| CFARC | Cloud-Free ARC |
| CFLOS | Cloud-Free Line-of-Sight |
| CMOS | Complementary Metal Oxide Semiconductor |
| CSI | Channel State Information |
| CW | Continuous-Wave |
| DAPPM | Differential Amplitude Pulse Position Modulation |
| DHPIM | Dual Header Pulse Interval Modulation |
| DOLCE | Deep Space Optical Link Communications Experiment |
| DPPM | Differential Pulse Position Modulation |
| DPSK | Differential Phase Shift Keying |
| DTN | Delay Tolerant Networking |
| DWDM | Dense Wavelength Division Multiplexing |
| EDRS | European Data Relay Satellite System |
| EGC | Equal Gain Combining |
| ESA | European Space Agency |
| ETS | Engineering Test Satellite |
| EW | Exponentiated Weibull |
| FELT | Free Space Experimental Laser Terminal |
| FIR | Far Infra-Red |
| FOV | Field-of-View |
| FPA | Focal Pixel Array |
| FSM | Fast Steering Mirrors |
| FSO | Free Space Optics |
| FXLMS | Filtered-X Least Mean Square |
| GA-ASI | General Atomics Aeronautical Systems, Inc |
| GEO | Geostationary Earth Orbit |
| GEOS-II | Geodetic Earth Orbiting Satellite-II |
| GOLD | Ground/Orbiter Lasercomm Demonstration |
| GOPEX | Galileo Optical Experiment |
| H-ARQ | Hybrid Automatic Repeat Request |
| HAP | High Altitude Platform |
| IEC | International Electrotechnical Commission |
| IF | Intermediate Frequency |
| IR | Infra-Red |
| ISR | Intelligence, Surveillance, and Reconnaissance |
| ITU | International Telecommunication Union |
| ITU | International Telecommunication Union |
| KIODO | KIrari's Optical Downlink to Oberpfaffenhofen |
| LCLS | Laser Cross Link System |
| LCRD | Laser Communications Relay Demonstration |
| LEO | Low Earth Orbit |
| LIDAR | Light Detection and Ranging |
| LIR | Long Infra-Red |
| LLCD | Lunar Laser Communication Demonstration |
| LMS | Least Means Squares |
| LO | Local Oscillator |
| LOLA | Airborne Laser Optical Link |
| LOS | Line-of-Sight |
| LQG | Linear-Quadratic Gaussian |
| MAC | Medium Access Control |
| MAP | Maximum-a-Posterior |
| MC-ARQ | Modified Cooperative-Diversity with Automatic Repeat Request |
| MEMS | Micro-Electro-Mechanical Systems |
| MIMO | Multiple Input Multiple Output |
| MIR | Mid Infra-Red |
| MISO | Multiple Input Single Output |
| ML | Maximum Likelihood |
| MLCD | Mars Laser Communications Demonstration |
| MLSD | Maximum-Likelihood Sequence Detection |
| MOLA | Mars Orbiter Laser Altimeter |
| MPE | Maximum Possible Exposure |
| MRC | Maximal Ratio Combining |
| MSIM | Multiple Sub-Sarrier Intensity Modulation |
| NIR | Near Iinfra-Red |
| OAF | Optical Amplify-and-Forward |
| OAM | Orbital Angular Momentum |
| OCD | Optical Communication Demonstrator |
| OFDM | Orthogonal Frequency Division Multiplexing |
| OFDM | Optical Frequency Division Multiplexing |
| OGS | Optical Ground Station |
| OICETS | Optical Inter-Orbit Communications Engineering Test Satellite |
| OIL | Optical Injection Locking |
| OIPLL | Optical Injection Phase-Lock Loop |
| OOK | On Off Keying |
| OPLL | Optical Phase-Lock Loop |
| OPPM | Overlapping Pulse Position Modulation |
| ORF | Optical Regenerate-and-Forward |
| OWC | Optical Wireless Communication |
| PAA | Point Ahead Angle |
| PAM | Pulse Amplitude Modulation |
| PAPR | Peak-to-Average Power Ratio |
| PDF | Probability Density Function |
| PIM | Pulse Interval Modulation |
| QAM | Quadrature Amplitude Modulation |
| QAPD | Quadrant Avalanche Photo-diodes |
| QoS | Quality of Service |
| RF | Radio Frequency |
| RIN | Relative Intensity Noise |
| RME | Relay Mirror Experiment |
| ROSA | RF Optical System for Aurora |
| RPA | Remotely Piloted Aircraft |
| RS | Reed-Solomon |
| RZ | Return-to-Zero |
| SAM | Spin Angular Momentum |
| SAR | Synthetic Aperture Radar |
| SC | Selection Combining |
| SC-BPSK | Sub-Carrier Binary Phase Shift Keying |
| SC-QPSK | Sub-Carrier Quadrature Phase Shift Keying |
| SILEX | Semiconductor Inter-Satellite Link Experiment |
| SIM | Sub-carrier Intensity Modulation |
| SIMO | Single Input Multiple Output |
| SIR | Short Infra-Red |
| SLC | Submarine Laser Communication |
| SLM | Spatial Light Modulator |
| SOLACOS | Solid State Laser Communications in Space |
| SOVA | Soft Output Viterbi Algorithm |
| SR-ARQ | Selective Repeat-Automatic Repeat Request |
| SROIL | Short Range Optical Inter-satellite Link |
| STR | Self-Tuning Regulator |
| STROPEX | Stratospheric Optical Payload Experiment |
| TCM | Trellis-Coded Modulation |
| TES | Tropospheric Emission Spectrometer |
| TFL | Tracking Feedback Loop |
| UAV | Unmanned Aerial Vehicle |
| VBLAST | Vertical Bell Laboratories Layered Space Time |
| VLSI | Very Large Scale Integration |
| WDM | Wavelength Division Multiplexing |

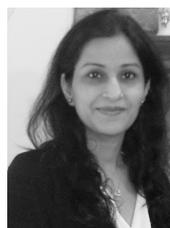

**Hemani Kaushal** received the Bachelor's degree in electronics and communication engineering from Punjab Technical University, India in 2001, Master's degree in electronics product design and technology from PEC University of Technology, Chandigarh, India in 2003 and Ph.D degree in electrical engineering from Indian Institute of Technology Delhi, India in 2012. She joined NorthCap University (former name ITM University), Gurgaon, India as Associate Professor in electrical, electronics and communication engineering department in 2012.

She has worked on various industry sponsored projects related to free space optical communication for Indian Space Research Organization (ISRO) and Aeronautical Development Agency (ADA), department of defense R&D, Bangalore, India. Her areas of research include wireless communication systems specifically in free-space, underwater, and indoor visible-light optical communications.

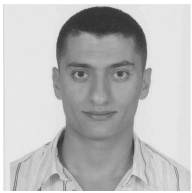

**Georges Kaddoum** (M'11) received the Bachelor's degree in electrical engineering from the École Nationale Supérieure de Techniques Avancées (ENSTA Bretagne), Brest, France, and the M.S. degree in telecommunications and signal processing (circuits, systems, and signal processing) from the Université de Bretagne Occidentale and Telecom Bretagne(ENSTB), Brest, in 2005 and the Ph.D. degree (with honors) in signal processing and telecommunications from the National Institute of Applied Sciences (INSA), University of Toulouse, Toulouse, France, in 2009. Since November 2013, he is an Assistant Professor of electrical engineering with the École de Technologie Supérieure (ETS), University of Quebec, Montréal, QC, Canada. In 2014, he was awarded the ETS Research Chair in physical-layer security for wireless networks. Since 2010, he has been a Scientific Consultant in the field of space and wireless telecommunications for several companies (Intelcan Techno-Systems, MDA Corporation, and Radio-IP companies). He has published over 80+ journal and conference papers and has two pending patents. His recent research activities cover wireless communication systems, chaotic modulations, secure transmissions, and space communications and navigation. Dr. Kaddoum received the Best Paper Award at the 2014 IEEE International Conference on Wireless and Mobile Computing, Networking, and Communications, with three coauthors, and the 2015 IEEE Transactions on Communications Top Reviewer Award.